\documentclass[traditabstract]{aa}
\usepackage{aalongtable}
\usepackage{natbib,twoopt}
 \bibpunct{(}{)}{;}{a}{}{,}    
\usepackage{graphics}
\usepackage{txfonts}
\usepackage{ulem}
\usepackage{times}
\usepackage{psfig}
\begin{document}
\title{Hunting for extremely metal-poor emission-line galaxies in the Sloan Digital
Sky Survey: MMT and 3.5m APO observations\thanks{Based on observations
with the Multiple Mirror telescope (MMT) and the 3.5m Apache Point Observatory 
(APO). The MMT is operated by 
the MMT Observatory (MMTO), a joint venture of the Smithsonian Institution  
and the University of Arizona. The Apache Point Observatory 
3.5-meter telescope is owned and operated by the Astrophysical Research 
Consortium.}$^,$\thanks{Figures 1 - 3 and Tables 2 - 8 are available
in electronic form at http://www.aanda.org.}}
\author{Y. I. \ Izotov \inst{1,2}    
\and T. X. \ Thuan \inst{3,4}
\and N. G. \ Guseva \inst{1,2}}
\offprints{Y. I. Izotov, izotov@mao.kiev.ua}
\institute{          Max-Planck-Institut f\"ur Radioastronomie, Auf dem H\"ugel 
                     69, 53121 Bonn, Germany
\and
                     Main Astronomical Observatory,
                     Ukrainian National Academy of Sciences,
                     Zabolotnoho 27, Kyiv 03680,  Ukraine
\and
                     Institut d'Astrophysique, Paris, 98 bis Boulevard Arago, 
                     75014, Paris, France
\and
                     Astronomy Department, University of Virginia, P.O. 
                     Box 400325, Charlottesville, VA 22904, USA
}
\date{Received \hskip 2cm; Accepted}

\abstract{
We present 6.5-meter MMT and 3.5m APO spectrophotometry
of 69 H {\sc ii} regions in 42 low-metallicity emission-line 
galaxies, selected from the Data Release 7 of the Sloan Digital
Sky Survey to have mostly [O {\sc iii}]$\lambda$4959/H$\beta$ $\la$ 1 and 
[N {\sc ii}]$\lambda$6583/H$\beta$ $\la$ 0.1. 
The electron temperature-sensitive emission line
[O {\sc iii}] $\lambda$4363 is detected in 53 H {\sc ii} regions
allowing a direct abundance determination. The oxygen
abundance in the remaining 16 H {\sc ii} regions is derived using 
a semi-empirical method.
The oxygen abundance of the galaxies in our sample ranges
from 12 + log O/H $\sim$ 7.1 to $\sim$ 7.9, with 14 
H {\sc ii} regions in 7 galaxies with 12 +log O/H $\leq$ 7.35. 
In 5 of the latter galaxies, the oxygen abundance is derived here 
for the first time.
Including other known extremely metal-deficient emission-line galaxies 
from the literature, e.g. SBS 0335$-$052W, SBS 0335$-$052E and I Zw 18, we 
have compiled a sample of the 17 most metal-deficient (with 
12 +log O/H $\leq$ 7.35)
emission-line galaxies known in 
the local universe. There appears to be a metallicity floor at 12 +log O/H 
$\sim$ 6.9, suggesting that the matter from which dwarf emission-line 
galaxies formed was pre-enriched to that level by e.g. Population III stars.}
\titlerunning{Hunting for extremely metal-poor emission-line galaxies}
\authorrunning{Y. I. Izotov et al.}
\maketitle

\section {Introduction}

Extremely metal-deficient (XMD) emission-line galaxies at low redshifts are the
most promising young galaxy candidates in the local Universe
\citep{G03,IT04b}. Those XMD objects are often defined as emission-line
galaxies with an oxygen abundance 
12 + log O/H $\leq$ 7.65 
\citep[e.g. ][]{K04a,K07,PM07,EC10}, a definition which we will adopt 
hereafter. 
These galaxies are important to identify 
because of several reasons. First, their nearly pristine interstellar 
medium (ISM) can shed light on the properties of the primordial ISM at 
the time of galaxy formation.     
There is now some evidence of a metallicity floor, i.e. that even 
the most metal-deficient star-forming galaxies in the local 
Universe formed from matter which was already pre-enriched by a previous star 
formation episode, e.g. by Population III stars
\citep{T05}. A similar metallicity floor appears also to be present
in damped Ly$\alpha$ systems \citep{P03}. 
It is thus quite important to find as many emission-line XMD 
galaxies as possible, especially the ones 
with the lowest metallicities, to have 
enough statistics to assess the reality of such a metallicity floor, and in so 
doing, to establish firmly the level of that pre-enrichment.
Second, because they have not undergone much chemical evolution, 
these galaxies
are also the best objects for the determination of the primordial $^4$He
abundance $Y_p$ 
\citep[e.g. ][]{IT04a,ITS07}. 
A recent determination of  $Y_p$ \citep{IT10} appears to indicate
 that a small deviation from the standard 
Big Bang nucleosynthesis model may exist, suggesting new physics such as 
the existence of light sterile neutrinos. 
It is important to confirm this result by improving the determination of $Y_p$ 
with as many XMD emission-line galaxies as possible.  
Third, in the hierarchical picture of galaxy 
formation, large galaxies form from the assembly of small dwarf 
galaxies.  
While much progress has been made in finding large populations of galaxies at 
high redshifts \citep[$z\ga\,3$, ][]{steidel03}, 
truly young galaxies in the process of forming remain 
elusive in the distant universe. 
The spectra of those far-away galaxies generally show the presence of a 
substantial amount of heavy elements, 
indicating previous star formation and metal enrichment. 
Therefore, XMD emission-line dwarf galaxies  
 are possibly the closest approximations we can find of the elementary 
primordial units from which galaxies formed. 
Their relative proximity allows studies of their 
stellar, gas and dust content with a sensitivity, 
spectral and spatial resolution that faint distant high-redshift galaxies 
do not allow.

\setcounter{table}{0}

\begin{table*}
  \caption{General Characteristics of Galaxies\label{tab1}}
\begin{tabular}{lrrcrcll} \hline \hline
Name&R.A.(J2000.0)&Dec.(J2000.0)&
redshift&Distance&$B$, $g$&Other names&Reference$^{\rm a}$ \\
&&&&Mpc&&& \\ \hline
\multicolumn{7}{c}{a) 3.5m APO observations} \\ \hline
J0113$+$0052No.1 &01:13:40.45 &$+$00:52:39.14 & 0.00383 &15.80 &13.69$\pm$0.01 & UGC 772&2 \\ 
J0113$+$0052No.2 &01:13:40.45 &$+$00:52:39.14 & 0.00383 &15.80 &13.69$\pm$0.01 & UGC 772&2 \\ 
J0113$+$0052No.3 &01:13:40.45 &$+$00:52:39.14 & 0.00383 &15.80 &13.69$\pm$0.01 & UGC 772&2 \\ 
J0113$+$0052No.4 &01:13:40.45 &$+$00:52:39.14 & 0.00383 &15.80 &13.69$\pm$0.01 & UGC 772&2 \\ 
J0138$+$0020     &01:38:35.02 &$+$00:20:05.12 & 0.01696 &67.84 &19.14$\pm$0.02 & &1 \\ %
J0834$+$5905No.1 &08:34:37.19 &$+$59:05:35.99 & 0.00477 &19.08 &19.87$\pm$0.02 & &1 \\ %
J0834$+$5905No.2 &08:34:37.19 &$+$59:05:35.99 & 0.00477 &19.08 &19.87$\pm$0.02 & &1 \\ 
J0843$+$4025     &08:43:37.99 &$+$40:25:46.96 & 0.00205 &10.17 &17.57$\pm$0.02 & &1 \\          %
J0851$+$8416     &08:51:03.55 &$+$84:16:13.33 & 0.00614 &29.70 &16.00$\pm$0.01 & UGC 4557 &1 \\ 
J0906$+$2528No.1 &09:06:00.92 &$+$25:28:11.33 & 0.00921 &39.50 &17.41$\pm$0.01 & &1 \\          %
J0906$+$2528No.2 &09:06:00.92 &$+$25:28:11.33 & 0.00921 &39.50 &17.41$\pm$0.01 & &1 \\          %
J0908$+$0517No.1 &09:08:36.53 &$+$05:17:26.95 & 0.00202 & 6.45 &16.96$\pm$0.24 & KKH 46 &3 \\   
J0908$+$0517No.2 &09:08:36.53 &$+$05:17:26.95 & 0.00208 & 6.45 &16.96$\pm$0.24 & KKH 46 &3 \\   
J0950$+$3127No.1 &09:50:19.49 &$+$31:27:22.24 & 0.00203 & 7.25 &15.41$\pm$0.01 & UGC 5272b &4 \\ 
J0950$+$3127No.2 &09:50:19.49 &$+$31:27:22.24 & 0.00201 & 7.25 &15.41$\pm$0.01 & UGC 5272b &4 \\ 
DDO68 No.2       &09:56:46.05 &$+$28:49:43.78 & 0.00203 & 5.85 &14.30$\pm$0.00 & UGC 534, J0956$+$2849No.2&5 \\ 
DDO68 No.3       &09:56:46.05 &$+$28:49:43.78 & 0.00196 & 5.85 &14.30$\pm$0.00 & UGC 534, J0956$+$2849No.3&5 \\ 
J1056$+$3608No.1 &10:56:40.35 &$+$36:08:27.94 & 0.00218 & 7.68 &18.54$\pm$0.02 & HS 1053$+$3624&1 \\ %
J1056$+$3608No.2 &10:56:40.35 &$+$36:08:27.94 & 0.00200 & 7.68 &18.54$\pm$0.02 & HS 1053$+$3624&1 \\ %
J1056$+$3608No.3 &10:56:40.35 &$+$36:08:27.94 & 0.00206 & 7.68 &18.54$\pm$0.02 & HS 1053$+$3624&1 \\ %
J1109$+$2007     &11:09:09.53 &$+$20:07:29.75 & 0.00373 &14.92 &17.20$\pm$0.01 & &1 \\ %
J1119$+$0935No.1 &11:19:28.09 &$+$09:35:44.28 & 0.00317 &12.10 &16.42$\pm$0.01 & &1 \\ 
J1119$+$0935No.2 &11:19:28.09 &$+$09:35:44.28 & 0.00343 &12.10 &16.42$\pm$0.01 & &1 \\ 
J1119$+$0935No.3 &11:19:28.09 &$+$09:35:44.28 & 0.00340 &12.10 &16.42$\pm$0.01 & &1 \\ 
J1121$+$3744No.1 &11:21:46.68 &$+$37:44:21.18 & 0.00679 &31.00 &17.87$\pm$0.01 & &1 \\ %
J1121$+$3744No.2 &11:21:46.68 &$+$37:44:21.18 & 0.00679 &31.00 &17.87$\pm$0.01 & &1 \\ %
J1146$+$4050     &11:46:37.60 &$+$40:50:36.68 & 0.00288 &15.90 &17.40$\pm$0.01 & &1 \\ %
J1226$-$0115No.1 &12:26:22.70 &$-$01:15:12.33 & 0.00659 &30.80 &16.26$\pm$0.01 & UM 501 &6 \\ 
J1226$-$0115No.2 &12:26:22.70 &$-$01:15:12.33 & 0.00659 &30.80 &16.26$\pm$0.01 & UM 501 &6 \\ 
J1235$+$2755No.1 &12:35:52.35 &$+$27:55:54.22 & 0.00291 &11.64 &16.72$\pm$0.14 &  &7 \\ 
J1235$+$2755No.2 &12:35:52.35 &$+$27:55:54.22 & 0.00280 &11.64 &16.72$\pm$0.14 &  &7 \\ 
J1235$+$2755No.3 &12:35:52.35 &$+$27:55:54.22 & 0.00280 &11.64 &16.72$\pm$0.14 &  &7 \\ 
J1241$-$0340     &12:41:59.35 &$-$03:40:02.42 & 0.00944 &40.10 &18.07$\pm$0.01 & &1 \\ %
J1257$+$3341No.1 &12:57:40.55 &$+$33:41:39.24 & 0.00299 &15.10 &17.43$\pm$0.01 & &1 \\ %
J1257$+$3341No.2 &12:57:40.55 &$+$33:41:39.24 & 0.00307 &15.10 &17.43$\pm$0.01 & &1 \\ %
J1257$+$3341No.3 &12:57:40.55 &$+$33:41:39.24 & 0.00308 &15.10 &17.43$\pm$0.01 & &1 \\ %
J1403$+$5804No.1 &14:03:21.46 &$+$58:04:03.44 & 0.00237 &14.26 &17.12$\pm$0.01 & &1 \\ %
J1403$+$5804No.2 &14:03:21.46 &$+$58:04:03.44 & 0.00242 &14.26 &17.12$\pm$0.01 & &1 \\ %
SBS1420+544      &14:22:38.80 &$+$54:14:10.00 & 0.00242 &88.90 &17.70$\pm$0.01 & J1422$+$5414 &1 \\ %
PHL293B          &22:30:36.79 &$-$00:06:36.96 & 0.00583 &24.40 &16.98$\pm$0.01 & J2230$-$0006 &1 \\ \hline %
\end{tabular}

$^{\rm a}$References on the photometric data. The $g$ magnitudes are from the
SDSS and the $B$ magnitudes are from other sources. The coding is 
as follows: (1) SDSS; (2) \citet{PT96}; 
(3) \citet{M09}; (4) \citet{K04}; (5) \citet{dV91};
(6) \citet{D05}; (7) \citet{OA95}.

  \end{table*}

\setcounter{table}{0}

\begin{table*}
  \caption{---{\sl Continued.}}
\begin{tabular}{lrrcrcll} \hline \hline
Name&R.A.(J2000.0)&Dec.(J2000.0)&
redshift&Distance&$B$, $g$&Other names&Reference$^{\rm a}$ \\
&&&&Mpc&&& \\ \hline
\multicolumn{7}{c}{b) low-resolution MMT observations} \\ \hline
J1016$+$3754     &10:16:24.53 &$+$37:54:45.97 & 0.00392 &19.90 &15.90$\pm$0.01 & HS 1013$+$3809 &1 \\ %
J1016$+$5823No.1 &10:16:53.11 &$+$58:23:40.33 & 0.00768 &35.40 &15.90$\pm$0.20 & UGC 5541 &5 \\ 
J1016$+$5823No.2 &10:16:53.11 &$+$58:23:40.33 & 0.00744 &35.40 &15.90$\pm$0.20 & UGC 5541 &5 \\ 
J1016$+$5823No.3 &10:16:53.11 &$+$58:23:40.33 & 0.00780 &35.40 &15.90$\pm$0.20 & UGC 5541 &5 \\ 
J1044$+$0353     &10:44:57.80 &$+$03:53:13.15 & 0.01296 &54.50 &17.48$\pm$0.06 & &1 \\ %
J1119$+$5130     &11:19:34.32 &$+$51:30:12.13 & 0.00473 &23.10 &16.94$\pm$0.00 & ARP'S GALAXY \\ 
J1132$+$5722No.1 &11:32:02.62 &$+$57:22:36.56 & 0.00507 &26.40 &16.19$\pm$0.01 & SBS 1129$+$576 &1 \\ %
J1132$+$5722No.2 &11:32:02.62 &$+$57:22:36.56 & 0.00534 &26.40 &16.19$\pm$0.01 & SBS 1129$+$576 &1 \\ %
J1132$+$5722No.3 &11:32:02.62 &$+$57:22:36.56 & 0.00521 &26.40 &16.19$\pm$0.01 & SBS 1129$+$576 &1 \\ %
J1154$+$4636     &11:54:41.22 &$+$46:36:36.35 & 0.00360 &19.30 &17.21$\pm$0.01 & CG 1473 &1 \\ %
J1215$+$5223     &12:15:46.62 &$+$52:23:13.85 & 0.00046 & 3.33 &15.17$\pm$0.00 & CGCG 269-049  &1\\ %
J1224$+$3724     &12:24:36.72 &$+$37:24:36.55 & 0.04040 &179.90&17.86$\pm$0.01 & CG 1022, HS 1222$+$3741 &1 \\ %
J1244$+$3212No.1 &12:44:05.98 &$+$32:12:32.33 & 0.00223 & 7.51 &10.10$\pm$0.01 & NGC 4656 &5 \\ 
J1244$+$3212No.2 &12:44:05.98 &$+$32:12:32.33 & 0.00206 & 7.51 &10.10$\pm$0.01 & NGC 4656 &5 \\ 
J1244$+$3212No.3 &12:44:05.98 &$+$32:12:32.33 & 0.00207 & 7.51 &10.10$\pm$0.01 & NGC 4656 &5 \\ 
J1248$+$4823     &12:48:27.79 &$+$48:23:03.29 & 0.02955 &124.60&18.09$\pm$0.01 & HS 1246$+$4839 &1 \\ %
J1327$+$4022     &13:27:23.29 &$+$40:22:04.15 & 0.01049 &48.00 &19.11$\pm$0.01 & &1 \\ %
J1331$+$4151     &13:31:26.90 &$+$41:51:48.29 & 0.01168 &52.60 &16.92$\pm$0.00 & &1 \\ %
J1355$+$4651     &13:55:25.64 &$+$46:51:51.34 & 0.02812 &118.30&19.26$\pm$0.01 & HS 1353$+$4706 &1 \\ %
J1608$+$3528     &16:08:10.36 &$+$35:28:09.30 & 0.03241 &138.90&18.75$\pm$0.02 & &1 \\ \hline %
\multicolumn{7}{c}{c) medium-resolution MMT observations} \\ \hline
J0810$+$1837     &08:10:30.66 &$+$18:37:04.22 & 0.00657 &21.70 &17.97$\pm$0.02 & &1 \\ %
J0833$+$2508No.1 &08:33:35.65 &$+$25:08:47.14 & 0.00759 &32.10 &18.00$\pm$0.01 & &1 \\ %
J0833$+$2508No.2 &08:33:35.65 &$+$25:08:47.14 & 0.00759 &32.10 &18.00$\pm$0.01 & &1 \\ %
J0834$+$5905     &08:34:37.19 &$+$59:05:35.99 & 0.00477 &19.08 &19.87$\pm$0.02 & &1 \\ %
J0931$+$2717No.1 &09:31:36.15 &$+$27:17:46.64 & 0.00507 &23.40 &17.71$\pm$0.01 & &1 \\ %
J0931$+$2717No.2 &09:31:36.15 &$+$27:17:46.64 & 0.00507 &23.40 &17.71$\pm$0.01 & &1 \\ %
DDO68 No.3       &09:56:46.05 &$+$28:49:43.78 & 0.00188 & 5.85 &14.30$\pm$0.00 & UGC 534, J0956$+$2849No.3 &1 \\ %
DDO68 No.4       &09:56:46.05 &$+$28:49:43.78 & 0.00185 & 5.85 &14.30$\pm$0.00 & UGC 534, J0956$+$2849No.4 &1 \\ %
J0959$+$4626     &09:59:05.76 &$+$46:26:50.49 & 0.00200 & 8.00 &17.80$\pm$0.01 & &1 \\ %
J1057$+$1358No.1 &10:57:38.15 &$+$13:58:44.53 & 0.00424 &20.20 &17.24$\pm$0.02 & &1 \\ 
J1057$+$1358No.2 &10:57:38.15 &$+$13:58:44.53 & 0.00424 &20.20 &17.24$\pm$0.02 & &1 \\ 
J1121$+$3744No.1 &11:21:46.68 &$+$37:44:21.18 & 0.00657 &31.00 &17.87$\pm$0.01 & &1 \\ %
J1121$+$3744No.2 &11:21:46.68 &$+$37:44:21.18 & 0.00657 &31.00 &17.87$\pm$0.01 & &1 \\ %
J1231$+$4205     &12:31:09.08 &$+$42:05:33.89 & 0.00190 & 8.19 &17.78$\pm$0.02 & &1 \\ \hline %
\end{tabular}

$^{\rm a}$References on the photometric data. The $g$ magnitudes are from the
SDSS and the $B$ magnitudes are from other sources. The coding is as follows: 
(1) SDSS; (2) \citet{PT96}; 
(3) \citet{M09}; (4) \citet{K04}; (5) \citet{dV91};
(6) \citet{D05}; (7) \citet{OA95}.

  \end{table*}

XMD emission-line galaxies are however very rare \citep[e.g. ][]{KO00}. 
Many surveys have been carried out 
to search for such galaxies without significant success. For more than three
decades,
one of the first blue compact dwarf (BCD) galaxies discovered, I Zw 18 
\citep{SS70}, continued to hold the record as the most metal-deficient 
emission-line galaxy known, with
an oxygen abundance 12 + log O/H = 7.17 $\pm$ 0.01 in its 
northwestern component 
and 7.22 $\pm$ 0.02 in its southeastern component \citep{TI05}. 
Only very recently, has I Zw 18 been displaced by the BCD 
SBS 0335--052W which was discovered by \citet{P97}. 
This galaxy, with an oxygen abundance 
12 + log O/H = 7.12 $\pm$ 0.03, is now the emission-line galaxy 
with the lowest metallicity known \citep{I05}. Recently, \citet{I09}
derived the oxygen abundance in four H {\sc ii} regions of SBS 0335--052W and
found that it varies from region to region in the range 6.86 -- 7.22.

Because of the scarcity of XMD emission-line galaxies such as I Zw 18 
and SBS 0335--052W, we stand a better
chance of finding them in very large spectroscopic surveys. 
One of the best surveys suitable for
such a search is the Sloan Digital Sky Survey (SDSS) \citep{Y00}. 
However, despite intensive studies of galaxies with a detected 
temperature-sensitive [O {\sc iii}]
$\lambda$4363 emission line in their spectra, no emission-line galaxy
with an oxygen abundance as low as that of I Zw 18 has been 
discovered in the SDSS Data Release 3 (DR3) and earlier releases. 
The lowest-metallicity emission-line galaxies found in these releases 
have oxygen abundances 12 + log O/H $>$ 7.4 \citep{K04a,K04b,I04,I06}. 
Only recently, have 
\citet{P05}, \citet{I06b}, \citet{IT07,IT09} and \citet{G07} 
found several galaxies with 12+log O/H $\leq$ 7.35 in 
later SDSS spectroscopic releases.

In order to find new candidates for XMD  
emission-line galaxies, we have carried out a systematic search 
for such objects in the SDSS Data Release 7 (DR7).
We have chosen them mainly
on the basis of the relative fluxes of selected  
emission lines, as described in \citet{I06b} and \citet{IT07}. 
All known XMD emission-line galaxies 
are characterised by relatively weak (compared to 
H$\beta$) [O {\sc ii}] $\lambda$3727, [O {\sc iii}] $\lambda$4959, 
$\lambda$5007 and [N {\sc ii}] $\lambda$6583 emission 
lines \citep[e.g. ][]{IT98a,IT98b,I05,P05,I06b}.
These spectral properties select out uniquely low-metallicity 
dwarfs since no other type of galaxy possesses them.
This technique appears to be more efficient in uncovering 
XMD galaxies with the lowest metallicities, those with 
12 + log O/H $\leq$ 7.35, as compared to the selection technique  
based on broad-band photometric colors used e.g. by
\citet{B08} who found only objects with 12 + log O/H $>$ 7.35.
In contrast to previous studies \citep{K04a,K04b,I04,I06} which focus 
exclusively 
on objects with a detected [O {\sc iii}] $\lambda$4363 emission line in the 
SDSS sprectra,
we have also considered objects which satisfy the criteria described above, 
but with spectra  
where [O {\sc iii}] $\lambda$4363 is weak or not detected.
Since the [O {\sc ii}] $\lambda$3727 line is out of the observed wavelength 
range in SDSS spectra of galaxies with a 
redshift $z$ lower than 0.02, we use the two following criteria 
to select candidate XMD emission-line galaxies:   
[O {\sc iii}]$\lambda$4959/H$\beta$ $\la$ 1 and 
[N {\sc ii}]$\lambda$6583/H$\beta$ $\la$ 0.1.
The efficiency of this technique to achieve our goals  
has been demonstrated by \citet{I06b} and \citet{IT07}. 

While the SDSS spectra allow us to select very low 
metallicity galaxies, we need additional spectral 
observations for the following reasons: 
1) spectra that go further into the blue 
wavelength range than SDSS spectra 
are required to detect the  [O {\sc ii}] $\lambda$3727 line. 
For a precise oxygen abundance determination, 
this line is needed  
to provide information on the singly ionised ionic population of oxygen. 
In principle, the [O {\sc ii}]
$\lambda$7320, 7330 emission lines can be used instead. However,
these lines are very weak or not detected in the SDSS spectra of 
low-metallicity candidates. 2) A better signal-to-noise ratio
spectrum may allow the detection of 
 a weak [O {\sc iii}] $\lambda$4363 emission 
line, which would permit a direct determination of the electron temperature.
3) Very often, low-metallicity candidates possess two or more 
H~{\sc ii} regions with different degrees of excitation. However, 
SDSS spectra are usually obtained for only one H {\sc ii} region,
chosen sometimes not in the most optimal way. 

For these reasons, we have obtained new spectroscopic observations
with the 3.5m APO telescope and the MMT 
of a sample of SDSS XMD emission-line galaxy candidates. 
In addition to the galaxies selected by using the criteria described above, we
have also selected a subsample of low-redshift ($z$$<$0.02) 
SDSS galaxies with high
excitation ([O {\sc iii}]$\lambda$4959/H$\beta$ $>$ 1) to improve
their abundance determination with MMT/Blue channel observations, because their
 [O {\sc ii}]$\lambda$3727 emission line is not included 
in the spectral range of their SDSS spectra. The motivation for 
observing  
this subsample is the fact that XMD BCDs like I Zw 18 and SBS 0335$-$052 
are also often characterised by high excitation.  
The observations and
data reduction are discussed in Section \ref{S2}. The element abundances are
derived in Section \ref{S3}. We consider in Section \ref{S4} the 
luminosity - metallicity relations for XMD emission-line galaxies and 
the issue of a metallicity floor, and 
in Section \ref{S5} their location in the emission-line diagnostic diagram. 
Our main findings are summarised in Section \ref{S6}.

\setcounter{table}{8}

\begin{table*}
  \caption{Abundances, extinction coefficients, H$\beta$ equivalent widths, 
and absolute magnitudes\label{tab9}} 
\begin{tabular}{lccccccrc} \hline \hline
  Name&12+logO/H$^a$&logN/O&logNe/O&logS/O&logAr/O&$C$(H$\beta$)&EW(H$\beta$)&$M_g$, $M_B$ \\ \hline
\multicolumn{9}{c}{a) 3.5m APO observations} \\ \hline
J0113$+$0052No.1&7.15$\pm$0.09(D)&  ...           &$-$0.72$\pm$0.21&  ...           &  ...           &0.065& 40.7&$-$17.62 \\
J0113$+$0052No.2&7.32$\pm$0.04(D)&$-$1.50$\pm$0.08&  ...           &  ...           &  ...           &0.035& 20.8&$-$17.62 \\
J0113$+$0052No.3&7.30$\pm$0.08(D)&  ...           &  ...           &  ...           &  ...           &0.115& 30.0&$-$17.62 \\
J0113$+$0052No.4&7.32$\pm$0.04(D)&$-$1.43$\pm$0.07&  ...           &  ...           &  ...           &0.045& 47.3&$-$17.62 \\
J0138$+$0020    &7.52$\pm$0.07(E)&  ...           &  ...           &  ...           &  ...           &0.210& 20.4&$-$15.15 \\
J0834$+$5905No.1&7.43$\pm$0.05(E)&$-$1.49$\pm$0.10&  ...           &  ...           &  ...           &0.285& 17.9&$-$12.03 \\
J0834$+$5905No.2&7.37$\pm$0.06(E)&  ...           &  ...           &  ...           &  ...           &0.135&  9.9&$-$12.03 \\
J0843$+$4025    &7.57$\pm$0.06(E)&$-$1.58$\pm$0.10&  ...           &  ...           &  ...           &0.000& 16.3&$-$12.47 \\
J0851$+$8416    &7.66$\pm$0.06(D)&$-$1.47$\pm$0.09&$-$0.93$\pm$0.14&  ...           &  ...           &0.135& 83.6&$-$16.75 \\
J0906$+$2528No.1&7.50$\pm$0.14(D)&$-$1.77$\pm$0.19&$-$0.93$\pm$0.43&  ...           &  ...           &0.210& 12.2&$-$16.22 \\
J0906$+$2528No.2&7.46$\pm$0.09(D)&$-$1.43$\pm$0.12&$-$0.81$\pm$0.22&  ...           &  ...           &0.240& 37.1&$-$16.22 \\
J0908$+$0517No.1&7.39$\pm$0.08(E)&$-$1.46$\pm$0.15&  ...           &  ...           &  ...           &0.410& 17.2&$-$12.73 \\
J0908$+$0517No.2&7.68$\pm$0.03(D)&$-$1.32$\pm$0.04&$-$0.76$\pm$0.05&$-$1.49$\pm$0.06&$-$2.17$\pm$0.04&0.040&197.4&$-$12.73 \\ %
J0950$+$3127No.1&7.41$\pm$0.07(E)&  ...           &  ...           &  ...           &  ...           &0.005& 14.5&$-$14.10 \\
J0950$+$3127No.2&7.45$\pm$0.17(D)&$-$1.62$\pm$0.25&  ...           &  ...           &$-$2.04$\pm$0.22&0.140&124.6&$-$14.10 \\
DDO68 No.2      &7.18$\pm$0.05(D)&  ...           &$-$0.79$\pm$0.10&  ...           &$-$2.28$\pm$0.07&0.030&191.6&$-$14.59 \\
DDO68 No.3      &7.08$\pm$0.09(D)&  ...           &$-$0.93$\pm$0.22&  ...           &  ...           &0.005& 85.3&$-$14.59 \\
J1056$+$3608No.1&7.16$\pm$0.07(D)&$-$1.45$\pm$0.25&$-$0.83$\pm$0.17&  ...           &  ...           &0.340&134.4&$-$11.21 \\
J1056$+$3608No.2&7.32$\pm$0.16(D)&$-$1.47$\pm$0.22&  ...           &  ...           &  ...           &0.000& 31.7&$-$11.21 \\
J1056$+$3608No.3&7.31$\pm$0.11(D)&$-$1.46$\pm$0.16&$-$0.70$\pm$0.32&  ...           &  ...           &0.000& 32.5&$-$11.21 \\
J1109$+$2007    &7.47$\pm$0.10(D)&$-$1.77$\pm$0.14&  ...           &  ...           &  ...           &0.350& 22.6&$-$14.67 \\
J1119$+$0935No.1&7.47$\pm$0.05(E)&$-$1.43$\pm$0.09&  ...           &  ...           &$-$2.49$\pm$0.13&0.055& 18.7&$-$14.53 \\
J1119$+$0935No.2&7.70$\pm$0.06(D)&$-$1.49$\pm$0.08&$-$0.87$\pm$0.13&  ...           &$-$2.47$\pm$0.07&0.225& 57.6&$-$14.53 \\
J1119$+$0935No.3&7.43$\pm$0.09(E)&$-$1.44$\pm$0.17&  ...           &  ...           &  ...           &0.280& 29.5&$-$14.53 \\
J1121$+$3744No.1&6.78$\pm$0.16(E)&  ...           &  ...           &  ...           &  ...           &0.000&  8.9&$-$14.59 \\
J1146$+$4050    &7.34$\pm$0.06(E)&$-$1.31$\pm$0.10&  ...           &  ...           &$-$2.21$\pm$0.18&0.110& 17.9&$-$13.92 \\ %
J1226$-$0115No.1&7.77$\pm$0.02(D)&$-$1.44$\pm$0.02&$-$0.78$\pm$0.03&$-$1.59$\pm$0.03&$-$2.20$\pm$0.03&0.130&158.5&$-$16.76 \\
J1226$-$0115No.2&7.81$\pm$0.02(D)&$-$1.37$\pm$0.02&$-$0.83$\pm$0.03&$-$1.55$\pm$0.03&$-$2.27$\pm$0.04&0.275&149.2&$-$16.76 \\
J1235$+$2755No.1&7.58$\pm$0.05(E)&$-$1.74$\pm$0.07&$-$0.82$\pm$0.14&  ...           &  ...           &0.105& 36.7&$-$13.99 \\ %
J1235$+$2755No.2&7.83$\pm$0.02(D)&$-$1.51$\pm$0.03&$-$0.76$\pm$0.05&$-$1.78$\pm$0.05&$-$2.42$\pm$0.03&0.075&129.8&$-$13.99 \\
J1235$+$2755No.3&7.77$\pm$0.03(D)&$-$1.52$\pm$0.04&$-$0.78$\pm$0.07&$-$1.80$\pm$0.05&$-$2.37$\pm$0.04&0.220&156.6&$-$13.99 \\
J1241$-$0340    &7.74$\pm$0.01(D)&$-$1.54$\pm$0.02&$-$0.69$\pm$0.03&$-$1.64$\pm$0.03&$-$2.27$\pm$0.02&0.310&125.3&$-$15.83 \\
J1257$+$3341No.1&7.47$\pm$0.11(D)&$-$1.44$\pm$0.15&$-$0.83$\pm$0.28&  ...           &$-$2.51$\pm$0.16&0.055& 75.7&$-$13.52 \\
J1257$+$3341No.2&7.62$\pm$0.07(E)&  ...           &$-$0.59$\pm$0.21&  ...           &  ...           &0.000& 22.8&$-$13.52 \\
J1257$+$3341No.3&7.48$\pm$0.09(D)&$-$1.53$\pm$0.13&$-$0.88$\pm$0.27&  ...           &$-$2.45$\pm$0.14&0.000& 90.6&$-$13.52 \\
J1403$+$5804No.1&7.36$\pm$0.06(E)&$-$1.49$\pm$0.12&  ...           &  ...           &  ...           &0.070&  9.1&$-$14.01 \\
J1403$+$5804No.2&7.40$\pm$0.05(D)&$-$1.64$\pm$0.08&$-$0.79$\pm$0.11&  ...           &$-$2.37$\pm$0.08&0.180& 46.4&$-$14.01 \\
SBS1420+544     &7.79$\pm$0.02(D)&$-$1.54$\pm$0.03&$-$0.69$\pm$0.03&$-$1.61$\pm$0.05&$-$2.21$\pm$0.06&0.105&201.8&$-$17.34 \\
PHL293B         &7.67$\pm$0.02(D)&$-$1.80$\pm$0.03&$-$0.70$\pm$0.03&$-$1.78$\pm$0.05&$-$2.40$\pm$0.05&0.115& 69.6&$-$15.29 \\ \hline
\end{tabular}


  \end{table*}

\setcounter{table}{8}

\begin{table*}
  \caption{---{\sl Continued.}} 
\begin{tabular}{lccccccrc} \hline \hline
  Name&12+logO/H$^a$&logN/O&logNe/O&logS/O&logAr/O&$C$(H$\beta$)&EW(H$\beta$)&$M_g$,$M_B$ \\ \hline
\multicolumn{9}{c}{b) low-resolution MMT observations} \\ \hline
J1016$+$3754    &7.64$\pm$0.01(D)&$-$1.55$\pm$0.03&$-$0.76$\pm$0.02&$-$1.52$\pm$0.03&$-$2.26$\pm$0.04&0.115& 96.3&$-$15.92 \\
J1016$+$5823No.1&7.74$\pm$0.01(D)&$-$1.71$\pm$0.02&$-$0.77$\pm$0.02&$-$1.68$\pm$0.04&$-$2.54$\pm$0.03&0.000&125.8&$-$17.29 \\
J1016$+$5823No.2&7.59$\pm$0.06(D)&$-$1.43$\pm$0.08&$-$0.88$\pm$0.14&  ...           &  ...           &0.440&102.6&$-$17.29 \\
J1016$+$5823No.3&7.62$\pm$0.01(D)&$-$1.72$\pm$0.03&$-$0.77$\pm$0.03&$-$1.84$\pm$0.07&$-$2.48$\pm$0.04&0.020&169.4&$-$17.29 \\ 
J1044$+$0353    &7.44$\pm$0.01(D)&$-$1.61$\pm$0.03&$-$0.75$\pm$0.02&$-$1.64$\pm$0.04&$-$2.27$\pm$0.04&0.150&257.5&$-$16.63 \\
J1119$+$5130    &7.51$\pm$0.02(D)&$-$1.64$\pm$0.04&$-$0.78$\pm$0.05&$-$1.69$\pm$0.11&$-$2.60$\pm$0.07&0.025& 30.4&$-$14.95 \\
J1132$+$5722No.1&7.50$\pm$0.03(D)&$-$1.56$\pm$0.04&$-$0.83$\pm$0.06&$-$1.57$\pm$0.06&$-$2.43$\pm$0.05&0.140& 81.8&$-$16.48 \\
J1132$+$5722No.2&7.74$\pm$0.03(D)&$-$1.38$\pm$0.04&$-$0.93$\pm$0.07&$-$1.64$\pm$0.08&$-$2.38$\pm$0.05&0.245& 41.0&$-$16.48 \\
J1132$+$5722No.3&7.68$\pm$0.05(D)&$-$1.44$\pm$0.09&$-$0.86$\pm$0.11&  ...           &  ...           &0.235& 22.3&$-$16.48 \\
J1154$+$4636    &7.75$\pm$0.02(D)&$-$1.37$\pm$0.03&$-$0.75$\pm$0.04&$-$1.56$\pm$0.05&$-$2.42$\pm$0.04&0.060& 89.7&$-$14.39 \\
J1215$+$5223    &7.53$\pm$0.02(D)&$-$1.56$\pm$0.04&$-$0.80$\pm$0.04&$-$1.63$\pm$0.08&$-$2.38$\pm$0.04&0.180& 84.1&$-$12.96 \\
J1224$+$3724    &7.78$\pm$0.01(D)&$-$1.39$\pm$0.02&$-$0.76$\pm$0.02&$-$1.65$\pm$0.04&$-$2.30$\pm$0.03&0.145&110.9&$-$18.83 \\
J1244$+$3212No.1&7.81$\pm$0.03(D)&$-$1.64$\pm$0.02&$-$0.80$\pm$0.03&$-$1.79$\pm$0.03&$-$2.44$\pm$0.02&0.085& 78.8&$-$19.67 \\
J1244$+$3212No.2&7.90$\pm$0.02(D)&$-$1.66$\pm$0.03&$-$0.81$\pm$0.04&$-$1.71$\pm$0.05&$-$2.44$\pm$0.03&0.235& 75.2&$-$19.67 \\
J1244$+$3212No.3&7.81$\pm$0.01(D)&$-$1.57$\pm$0.02&$-$0.84$\pm$0.03&$-$1.73$\pm$0.03&$-$2.39$\pm$0.02&0.090&151.7&$-$19.67 \\
J1248$+$4823    &7.80$\pm$0.01(D)&$-$1.54$\pm$0.03&$-$0.76$\pm$0.02&$-$1.59$\pm$0.03&$-$2.19$\pm$0.03&0.125&114.5&$-$17.74 \\
J1327$+$4022    &7.70$\pm$0.01(D)&$-$1.37$\pm$0.03&$-$0.77$\pm$0.02&$-$1.37$\pm$0.04&$-$2.07$\pm$0.04&0.060&220.9&$-$14.47 \\ %
J1331$+$4151    &7.70$\pm$0.01(D)&$-$1.55$\pm$0.02&$-$0.76$\pm$0.02&$-$1.52$\pm$0.02&$-$2.20$\pm$0.02&0.135&177.6&$-$17.07 \\
J1355$+$4651    &7.58$\pm$0.01(D)&$-$1.44$\pm$0.07&$-$0.76$\pm$0.02&$-$1.58$\pm$0.07&$-$2.08$\pm$0.06&0.070&353.5&$-$16.30 \\
J1608$+$3528    &7.79$\pm$0.02(D)&$-$1.70$\pm$0.09&$-$0.68$\pm$0.02&  ...           &$-$2.20$\pm$0.09&0.440&367.5&$-$18.22 \\ \hline
\multicolumn{9}{c}{c) high-resolution MMT observations} \\ \hline
J0810$+$1837    &7.83$\pm$0.20(D)&  ...           &$-$0.84$\pm$0.60&  ...           &  ...           &0.000&  8.4&$-$13.71 \\
J0833$+$2508No.1&7.69$\pm$0.04(D)&  ...           &$-$0.89$\pm$0.10&  ...           &  ...           &0.000& 45.4&$-$14.67 \\
J0833$+$2508No.2&7.43$\pm$0.13(D)&  ...           &$-$0.78$\pm$0.39&  ...           &  ...           &0.095& 24.7&$-$14.67 \\
J0834$+$5905    &7.24$\pm$0.08(D)&  ...           &$-$0.89$\pm$0.23&  ...           &  ...           &0.010& 28.3&$-$11.56 \\ %
J0931$+$2717No.1&7.45$\pm$0.06(E)&  ...           &$-$1.01$\pm$0.22&  ...           &  ...           &0.000& 16.5&$-$14.14 \\ %
J0931$+$2717No.2&7.52$\pm$0.06(E)&  ...           &$-$1.01$\pm$0.22&  ...           &  ...           &0.000& 18.2&$-$14.14 \\ %
DDO68 No.3      &7.15$\pm$0.04(D)&  ...           &$-$0.81$\pm$0.07&  ...           &  ...           &0.000& 15.7&$-$14.54 \\ %
DDO68 No.4      &7.16$\pm$0.09(D)&  ...           &$-$0.92$\pm$0.26&  ...           &  ...           &0.000& 36.0&$-$14.54 \\ %
J0959$+$4626    &7.50$\pm$0.05(E)&  ...           &$-$0.91$\pm$0.22&  ...           &  ...           &0.265& 57.0&$-$12.47 \\ %
J1057$+$1358No.1&7.18$\pm$0.07(E)&  ...           &  ...           &  ...           &  ...           &0.000&  8.7&$-$14.29 \\
J1121$+$3744No.1&7.12$\pm$0.06(E)&  ...           &  ...           &  ...           &  ...           &0.000&  8.4&$-$14.59 \\
J1231$+$4205    &7.62$\pm$0.04(D)&  ...           &$-$0.87$\pm$0.10&  ...           &  ...           &0.190& 65.7&$-$12.33 \\ \hline
\end{tabular}

$^a$Oxygen abundances derived by direct (D) and semi-empirical (E) methods. 

  \end{table*}

\section {Observations and data reduction \label{S2}}

\subsection{APO observations}

New optical spectra were obtained for a subsample of galaxies  
using the 3.5 m APO telescope in the course of several nights 
during the February 
2008 - February 2009 period. The general characteristics of these galaxies, 
such as redshift, distance, apparent $B$ magnitude (if available in
the literature) or SDSS $g$ magnitude and other names
are shown in Table \ref{tab1}. 
The Journal of observations is shown in Table \ref{tab2} (available only in
the electronic edition).
The 3.5 m APO observations were made with the Dual Imaging Spectrograph (DIS) 
in the
both the blue and red wavelength ranges. In the blue range, we use the 
B400 grating with a linear dispersion of 
1.83 $\AA$/pix and a central wavelength of 4400$\AA$,  while in the red range
we use the R300 grating with a linear dispersion 2.31 $\AA$/pix and 
a central wavelength of 7500$\AA$.
The above instrumental set-up gave a spatial scale along the slit of 0\farcs4
pixel$^{-1}$, a spectral range $\sim$3600 -- 9600$\AA$\ and a spectral 
resolution of 7$\AA$\ (FWHM).
The slit orientation, the total exposure time and the airmass during
observations are shown in Table \ref{tab2}. 
Several Kitt Peak IRS spectroscopic standard stars were observed for flux
calibration. Spectra of He-Ne-Ar comparison arcs were obtained 
at the beginning or the end of each night for wavelength calibration. 

\setcounter{figure}{3}


\begin{figure}
\hspace*{0.0cm}\psfig{figure=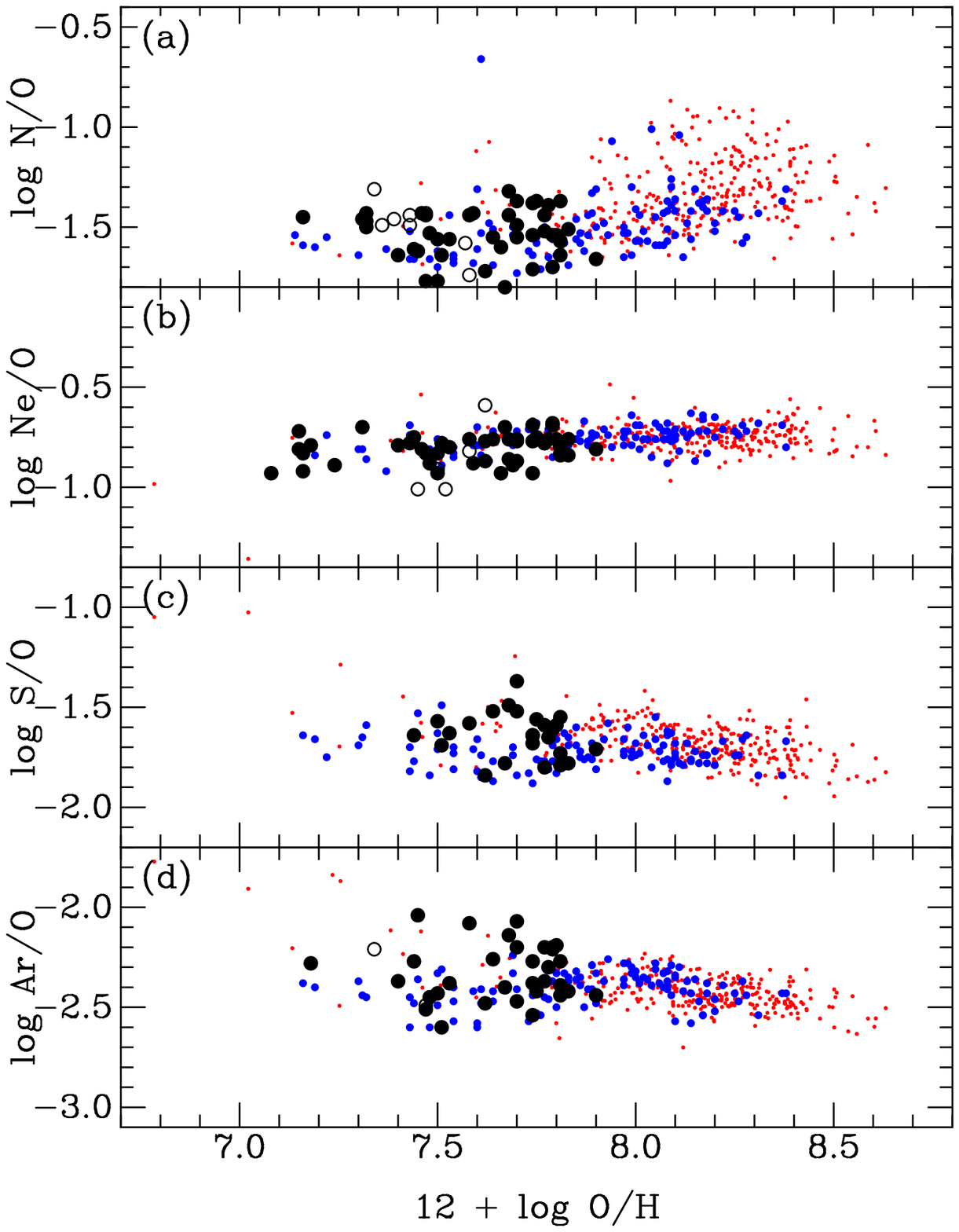,angle=0,width=8.8cm,clip=}
\caption{Abundance ratios vs oxygen abundance. The emission-line galaxies 
from this paper are shown by large circles, the filled circles representing 
galaxies with element abundances derived by the direct method, and the open 
ones galaxies with element
abundances derived by the semi-empirical method; 
small filled circles are XMD emission-line galaxies from \citet{IT04a}, 
and dots are SDSS galaxies from \citet{I06}.}
\label{fig4}
\end{figure}

\subsection{MMT observations}

The MMT spectrophotometric observations were obtained on the nights of 
2008 March 28 -- 29, and 2010 April 13.
The galaxies are listed in Table \ref{tab1} separately for low- and medium-
spectral resolution observations.
   All observations have been made with the Blue Channel of the MMT 
spectrograph.  The log of the observations is given in Table \ref{tab2}.
We used a 300 grooves/mm grating in first order for the low-resolution
observations and a 800 grooves/mm grating in first order for the 
medium-resolution observations. The above instrumental set-ups gave a spatial 
scale along the slit of 0\farcs6 pixel$^{-1}$, and respectively 
a spectral range of 
3600--7500$\AA$\ and a spectral resolution of 
$\sim$ 7$\AA$\ (FWHM) 
for the low-resolution observations, and of 3200--5200$\AA$\  
and $\sim$ 3$\AA$\ (FWHM)  
for the medium-resolution ones. 
The total exposure times, orientations of the slit and airmasses for 
each MMT observation are given in Table \ref{tab2}.
     In 2008, the Kitt Peak IRS spectroscopic standard stars BD 33d2642 
and HZ 44 were observed for flux calibration during the first night, while
BD 33d2642, BD 75d325 and Feige 34 were observed during the second
night. In 2010, BD 33d2642, BD 75d325 and Feige 34 were observed. 
Spectra of He--Ne--Ar comparison arcs were obtained
before and after each observation to calibrate the wavelength scale.

\subsection{Results}

Several galaxies listed in Table \ref{tab1} have been observed previously 
by other authors. 
To improve their abundance determination, we have reobserved  
J0113$+$0052, DDO 68, SBS 1116+517, SBS 1129+576, SBS 1420$+$544 and PHL 293B.
Two galaxies, J1121+3744 and J0834+5905, were observed twice, 
both with the 3.5m APO and the MMT.

The two-dimensional spectra were bias subtracted and flat-field corrected
using IRAF\footnote{IRAF is distributed by National Optical Astronomical 
Observatory, which is operated by the Association of Universities for 
Research in Astronomy, Inc., under cooperative agreement with the National 
Science Foundation.}. We then use the IRAF
software routines IDENTIFY, REIDENTIFY, FITCOORD, TRANSFORM to 
perform wavelength
calibration and correct for distortion and tilt for each frame. 
One-dimensional spectra were then extracted from each frame using the APALL 
routine. 
The sensitivity curve was obtained by 
fitting with a high-order polynomial the observed spectral energy 
distribution of standard stars. Then sensitivity curves obtained from
observations of different stars during the same night were averaged.

The 3.5m APO,
the low- and medium-resolution MMT  H {\sc ii} region spectra are 
shown respectively 
in Figures \ref{fig1}, \ref{fig2}
and \ref{fig3}. They are available only in the electronic 
online version. These spectra have all been reduced to zero redshift. 

   The extinction-corrected line fluxes $I$($\lambda$), normalised to 
$I$(H$\beta$) and multiplied by a factor of 100, and
their errors, derived from the 3.5m APO, 
MMT low-resolution and MMT 
medium-resolution observations are 
given in Tables \ref{tab3}, \ref{tab4} and \ref{tab5}, respectively. They are
available only in the electronic version on line. 
The data were obtained using the IRAF SPLOT routine.
The line flux errors listed include statistical errors derived with
SPLOT from non-flux calibrated spectra, in addition to errors introduced
in the standard star absolute flux calibration, which we set to 1\% of the
line fluxes. These errors will be later propagated into the calculation
of abundance errors.
The line fluxes were corrected for both reddening \citep[using 
the extinction curve of ][]{W58} 
and underlying hydrogen stellar absorption derived simultaneously by an 
iterative procedure as described in \citet{ITL94}. 
The extinction coefficient is defined as $C$(H$\beta$) = 1.47$E(B-V)$,
where $E(B-V)$ = $A(V)$/3.2 and $A(V)$ is the extinction in the $V$ band
\citep{A84}.
Equivalent widths EW(H$\beta$), extinction
coefficients $C$(H$\beta$), and equivalent widths EW(abs) of the hydrogen
absorption stellar lines are also given in Tables \ref{tab3} - \ref{tab5}, 
along with the uncorrected H$\beta$ fluxes.

\section {Physical conditions and element abundances \label{S3}}

   To determine element abundances, we generally follow  
the procedures of \citet{ITL94,ITL97} and \citet{TIL95}.
We adopt a two-zone photoionised H {\sc ii}
region model: a high-ionisation zone with temperature $T_e$(O {\sc iii}), 
where [O {\sc iii}] and [Ne {\sc iii}] lines originate, and a 
low-ionisation zone with temperature $T_e$(O {\sc ii}), where [O {\sc ii}], 
[N {\sc ii}] and [S {\sc ii}] lines originate. 
As for the [S {\sc iii}] and [Ar {\sc iii}] lines they originate in the 
intermediate zone between the high and low-ionisation regions. 
In the H {\sc ii} regions with a detected [O {\sc iii}] $\lambda$4363
emission line, the temperature $T_e$(O {\sc iii}) is calculated using the 
``direct'' method based on the 
[O {\sc iii}] $\lambda$4363/($\lambda$4959+$\lambda$5007) line ratio.
In H {\sc ii} regions where the [O {\sc iii}] $\lambda$4363 emission line 
is not detected,
we used a semi-empirical method described by \citet{IT07} to derive
$T_e$(O {\sc iii}).

For $T_e$(O {\sc ii}), we use
the relation between the electron temperatures $T_e$(O {\sc iii}) and
$T_e$(O {\sc ii}) obtained by \citet{I06} from
the H {\sc ii} photoionisation models of \citet{SI03}. These are based on
more recent stellar atmosphere models and improved
atomic data as compared to the \citet{S90} models. 
Ionic and total heavy element abundances are derived 
using expressions for ionic abundances and ionisation correction 
factors obtained by \citet{I06}.
The element abundances are given in 
Tables \ref{tab6}, \ref{tab7} and \ref{tab8} (available only in the electronic 
version on line) along with the adopted electron temperatures for
different ions. 

Table \ref{tab9} summarises the abundance determinations for all H {\sc ii}
regions, where we show the oxygen abundance 12 + log O/H, the logarithms
of abundance ratios log N/O, log Ne/O, log S/O and log Ar/O, the extinction
coefficient $C$(H$\beta$), the equivalent width of the H$\beta$ emission line,
and the galaxy's absolute magnitude in the SDSS $g$ band or in the $B$ band
whenever available from the literature. 
Examination of this Table shows that our sample consists of H {\sc ii}
regions spanning a wide range of oxygen abundance 12 + log O/H = 7.1 -- 7.9.
That range is clearly on the low metallicity side of the roughly Gaussian 
metallicity distribution of  
BCDs which peaks at 12 + log O/H $\sim$ 8.1 \citep{T08}. 
For comparison, the 12 + log O/H of the Small and Large Magellanic Clouds 
and the Sun are respectively 8.03, 8.35 \citep{RD92} and 8.69 \citep{A09}. 
One of the 
H {\sc ii} regions, J1121+3744 No.1, has an abnormally low oxygen abundance.  
We derived 12 + log O/H = 6.78$\pm$0.16
 from its 3.5m APO spectrum.
However, because of its faintness, the APO spectrum is
noisy (Fig. \ref{fig1}), and we have reobserved J1121+3744 No.1 with the MMT.   
The oxygen abundance obtained
from the MMT spectrum (Fig. \ref{fig3}) is significantly higher,
12 + log O/H = 7.12. However, it is still among the lowest known. Note, however,
that no [O {\sc iii}] $\lambda$4363 emission line is detected, therefore
its abundance has been derived using the semi-empirical method, 
making it less certain. 
The present sample contains 7 galaxies with 12 + log O/H $\leq$ 7.35. 
For two galaxies, J0113+0052 and DDO 68, we confirm the very low oxygen
abundances first found by \citet{I06b} and \citet{P05}, respectively.
As for the remaining five XMD galaxies, J0834+5905, J1056+3608, 
J1057+1358, J1121+3744 and J1146+4050, their oxygen abundances are derived for 
the first time in the 
present paper. 

In Table \ref{tab10} we show, in order of increasing oxygen abundance, 
the list of the 17 lowest metallicity XMD emission-line
galaxies known so far, including the galaxies studied in this paper. 
It is seen from Table \ref{tab10} that no H {\sc ii} region with an
oxygen abundance 12 + logO/H $<$ 7.1 has been found, except for
three H {\sc ii} regions (out of four observed) in SBS 0335--052W. 
This supports the idea
discussed by, e.g., 
\citet{T05} that the matter from which dwarf emission-line galaxies
formed was pre-enriched to a level 12 + log O/H $\ga$ 6.9 \citep[or $\sim$
2\% of the abundance 12 + log O/H = 8.69 of the Sun, ][]{A09}.
Based on {\sl FUSE} spectroscopic data, 
\citet{T05} showed that  
BCDs spanning a wide range in ionised gas metallicities all have  
H {\sc i} envelopes with about the same 
neutral gas metallicity of $\sim$7.0. This is also the metallicity  
found in Ly$\alpha$ absorbers \citep{P03}. Taken together, the available data 
suggest that there may have been 
previous enrichment of the primordial neutral gas to a common 
metallicity level 12 + log O/H $\sim$ 6.9, possibly by Population III stars.

As for other heavy elements, we show in Fig. \ref{fig4} by large circles
the location of our sample galaxies in the log N/O, log Ne/O, log S/O and
log Ar/O vs. oxygen abundance diagrams. The filled large 
circles represent galaxies 
with element abundances determined by the direct method while the 
open large circles show those with element abundances determined by the 
semi-empirical method. Examination of Fig. \ref{fig4} shows that there is 
no systematic difference between the filled and open large circles.    
For comparison, smaller filled 
circles show the galaxies from the sample of \citet{IT04a} used for
the primordial He abundance determination, and dots show the emission-line
galaxies from the SDSS DR3 \citep{I06}. In general, the locations of our
galaxies are consistent with those obtained for other emission-line galaxies.
However, since many of the galaxies studied in this paper are faint and with
weak emission lines in their spectra, they show a larger scatter in
Figure \ref{fig4} as compared to galaxies from the comparison sample.
Higher signal-to-noise ratio observations with large telescopes are
needed to improve the abundance determination in these galaxies.

\setcounter{table}{9}

\begin{table}
  \caption{List of the most metal-deficient emission-line
galaxies known \label{tab10}}
\begin{tabular}{lcc} \hline \hline
Name&12+logO/H$^a$&Reference$^b$ \\ \hline
SBS 0335$-$052W    & 6.86 -- 7.22(D)  & 3   \\
SBS 0335$-$052E    & 7.11 -- 7.32(D)  & 3   \\
J092609.45+334304.1&7.12(D)           &10   \\
J1121$+$3744       & 7.12(E)          & 1   \\
DDO68              & 7.14 -- 7.16(D)  & 1,4,6 \\
J0113$+$0052       & 7.15 -- 7.32(D)  & 1,2 \\
J1056$+$3608       & 7.16 -- 7.32(D)  & 1   \\
I Zw 18            & 7.17 -- 7.22(D)  & 5   \\
J1057$+$1358       & 7.18(E)          & 1   \\
J0834$+$5905       & 7.24(D)          & 1   \\
J2104$-$0035       & 7.26(D)          & 2   \\
J0254$+$0035       & 7.28(D)          & 4   \\
J0812$+$4836       & 7.28(D)          & 4   \\
Leo A              & 7.30(D)          & 7   \\
UGCA 292           & 7.30(D)          & 8   \\
J1414$-$0208       & 7.32(D)          & 9   \\
J1146$+$4050       & 7.34(E)          & 1   \\ \hline
\end{tabular}

$^a$Oxygen abundances derived by direct (D) and semi-empirical (E) methods.

$^b$(1) this paper; (2) \citet{I06b}; (3) \citet{I09}; 
(4) \citet{IT07}; (5) \citet{TI05}; (6) \citet{IT09}; (7) \citet{S89}, 
\citet{V06}; (8) \citet{V00}; (9) \citet{G07}; (10) \citet{P10}.
  \end{table}

\section{XMD emission-line galaxies and the luminosity-metallicity
relation \label{S4}}

  It was established more than 30 years ago that low-luminosity dwarf galaxies
have systematically lower metallicities as compared to more luminous galaxies 
\citep{L79,S89,RM95}. The study of this dependence  
sheds light on the physics of galaxy and star formation. The relation  
may also be potentially useful in searching for XMD  emission-line 
galaxies, in addition to our selection criterion based on the fluxes of
[O {\sc iii}] $\lambda$4959, 5007 and [N {\sc ii}] $\lambda$6584 emission 
lines relative to H$\beta$.

\setcounter{figure}{4}


\begin{figure*}
\hspace*{0.0cm}\psfig{figure=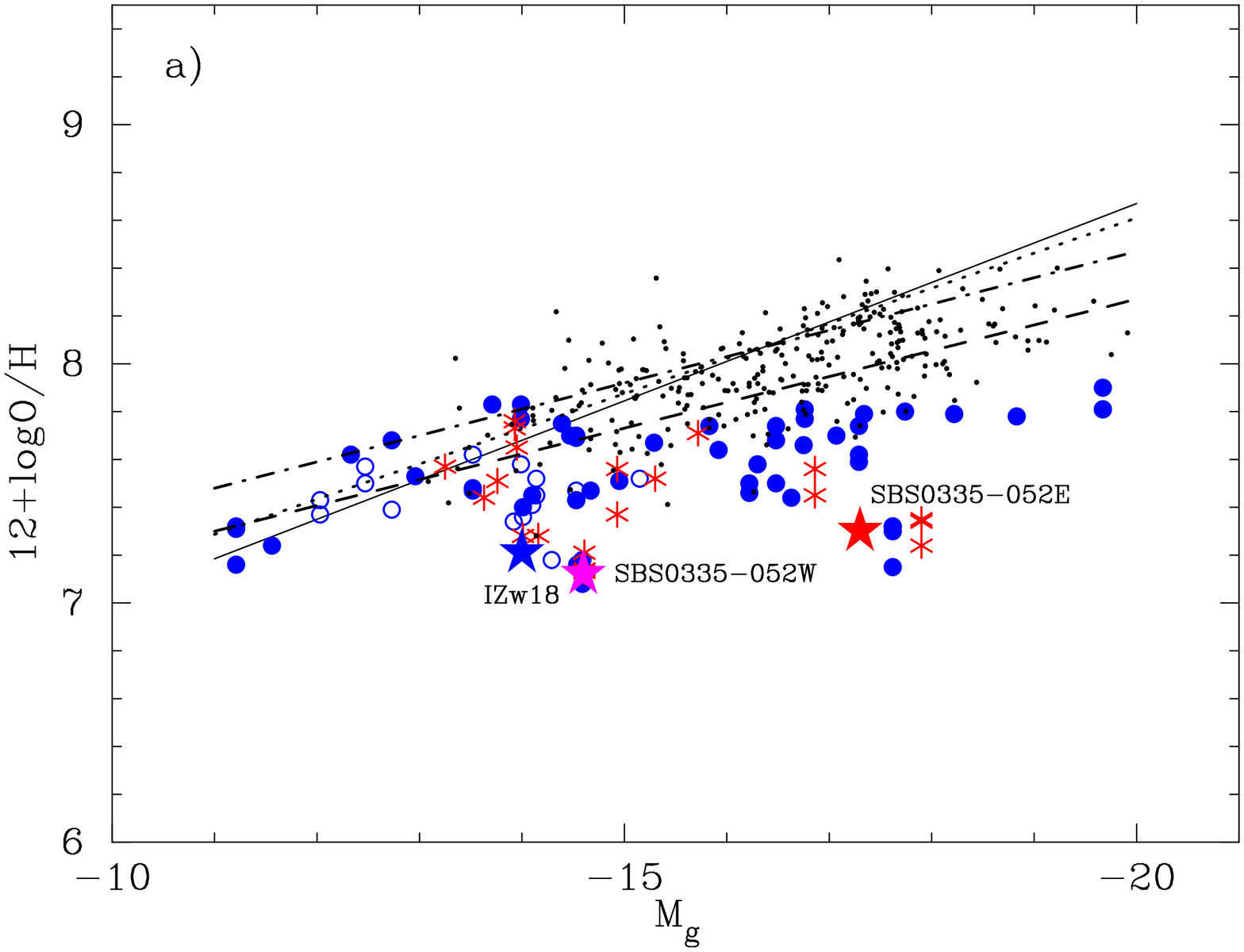,angle=-90,width=8.8cm,clip=}
\hspace*{0.3cm}\psfig{figure=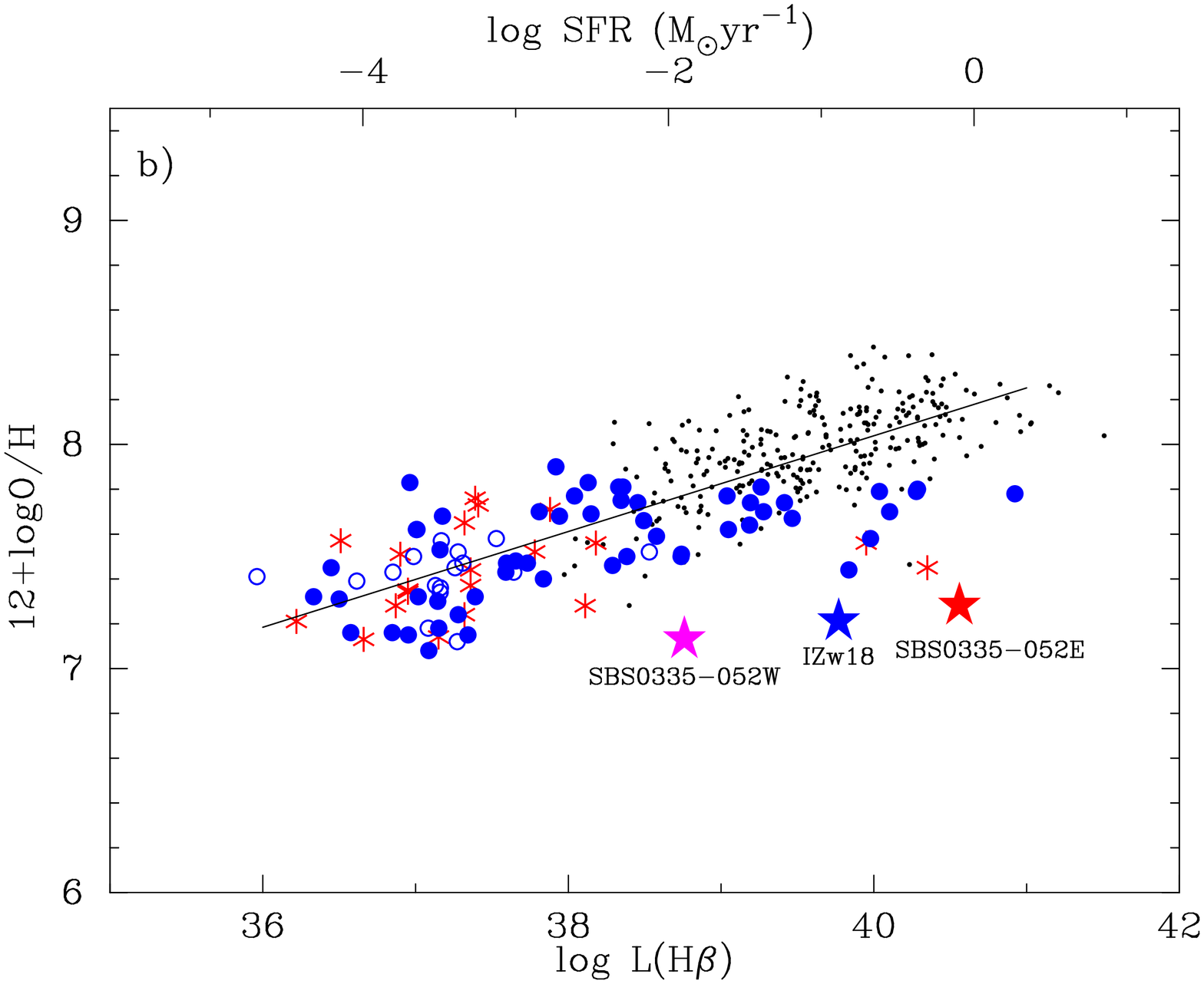,angle=-90,width=8.8cm,clip=}
\caption{(a) Absolute $g$ magnitude - oxygen abundance relation.
The emission-line galaxies 
from this paper are shown by large circles, the filled circles representing 
galaxies with element
abundances derived by the direct method, and the open ones galaxies with 
element abundances derived by the semi-empirical method; 
the XMD emission-line galaxies
of \citet{IT07} are shown by red asterisks, and the SDSS galaxies of 
\citet{I06} by black dots. The dashed line is the linear regression
to all these data. The solid line
is the relation derived by \citet{S89}, the dotted line is  
the one derived by \citet{RM95}, and the dot-dashed line is the one
derived by \citet{B12}. For comparison, the most-metal deficient BCDs 
I Zw 18, SBS 0335--052W and SBS 0335--052E are shown by blue, magenta and red
stars, respectively, and labelled. (b) H$\beta$ luminosity - oxygen abundance 
relation. The symbols are the same as in (a). The solid line is the linear 
regression to all data, excluding I Zw 18, SBS 0335--052W and SBS 0335--052E.}
\label{fig5}
\end{figure*}

As a measure of metallicity and luminosity of each galaxy, we adopt 
respectively
the oxygen abundance 12 + log O/H and the absolute magnitude $M_g$ in the $g$
band or $M_B$ in the $B$ band. We will call their dependence on one 
another the ``luminosity - metallicity'' or ``$L-Z$'' relation. 
Because \citet{P08} have shown that, for 
emission-line galaxies, the $B$--$g$ color index is $<$ 0.1 mag, 
we have not transformed $B$ magnitudes to $g$ magnitudes 
and have used either one, whenever
available in the literature. 
For the galaxy distances, we have selected the ones from the NASA/IPAC 
Extragalactic Database (NED)\footnote{This research has made use 
of the NASA/IPAC Extragalactic Database (NED) which is operated by the Jet 
Propulsion Laboratory, California Institute of Technology, under contract 
with the National Aeronautics and Space Administration.} which are obtained 
from the radial velocities corrected for Virgo infall only. Distances and
apparent $g$ or $B$ magnitudes are given in Table \ref{tab1} for all our 
galaxies.
In the cases where several  H {\sc ii} regions were observed within the same 
galaxy, the same distance and $g$ or $B$ magnitude of the whole galaxy 
were assigned to each H {\sc ii} region. The derived absolute
magnitudes $M_g$ or $M_B$ for all the galaxies in our sample are shown in
Table \ref{tab9}. They were obtained from the apparent magnitudes
in Table \ref{tab1} and 
corrected for extinction as defined by $C$(H$\beta$) in Table \ref{tab9}
for a single star-forming region, or by the average value of $C$(H$\beta$) 
in the case of multiple knots of star formation.

The luminosity-metallicity diagram for the present 
sample (Table \ref{tab1}) is shown in Figure \ref{fig5}a 
by blue large filled and open circles. The XMD emission-line 
galaxies discussed by \citet{IT07}, also selected from the SDSS with the same
selection criteria, are shown by red asterisks. For comparison, the 
dots show the large sample of emission-line galaxies studied by
\citet{I06}. The latter sample includes all emission-line 
star-forming galaxies from the SDSS DR3 with a 
reliably detected [O {\sc iii}] $\lambda$4363 emission
line (its flux is at least 4 times the flux error).
The distances to the SDSS DR3
galaxies are derived from their radial velocities, adopting a Hubble constant
$H_0$ = 75 km s$^{-1}$ Mpc$^{-1}$. All nearby objects with $z$ $<$ 0.003,
and hence with a less certain distance, are excluded from the SDSS DR3 sample.
We have also excluded all SDSS galaxies with $z$ $>$ 0.03, thus
restricting our sample to dwarf galaxies.
Totally, the SDSS DR3 sample in Fig. \ref{fig5}a contains 272 galaxies.
We have also shown and labelled the most-metal deficient BCDs known, 
SBS 0335--052W, SBS 0335--052E and I Zw 18, by blue, magenta and red
stars, respectively.
We show respectively by the solid, dotted, and dot-dashed lines the $L-Z$ 
relations obtained by \citet{S89}, \citet{RM95}, and \citet{B12}
for nearby dwarf irregular galaxies. 

It can be seen from Fig. \ref{fig5}a that the 
SDSS DR3 galaxies with $M_g$$>$$-$18 mag
are slightly offset to brighter magnitudes as compared to the $L-Z$ relations 
of \citet{S89}, \citet{RM95}, and \citet{B12}. 
Fitting the data set composed of the SDSS DR3 galaxies (black dots) and the
emission-line galaxies from  \citet{IT07} (red asterisks) and from 
the present paper (blue filled and open circles)   
by a linear regression gives: 
\begin{equation}
{\rm 12+logO/H} = -0.108\times M_g+6.113. \label{eq1}
\end{equation}
This linear regression is shown in Fig. \ref{fig5}a by a dashed line. It 
has a slope shallower than the ones of $-$0.165 from \citet{S89}
and of $-$0.147 from \citet{RM95}. On the other hand, the slope of
our relation (Eq. \ref{eq1}) is the same as that by \citet{B12}. However,
the relation defined by Eq. \ref{eq1} is shifted by $\sim$ 0.2 dex to lower 
oxygen abundances indicating that we find more metal-poor galaxies with our 
selection criteria.  
The spread of the data points about the fit is partially due to 
observational uncertainties, but also to the way the samples are selected.  
The SDSS DR3 galaxies have been selected to have a detectable 
[O {\sc iii}] $\lambda$4363 emission line, and hence to have 
strong star formation, in contrast to the more quiescent dwarf 
irregular galaxies studied by \citet{S89} and \citet{RM95}. 
The shallower slope of the relation defined by Eq. \ref{eq1} 
as compared to the relations by \citet{S89} and \citet{RM95} is 
due to the offset, at a given metallicity, 
to brighter magnitudes of the SDSS DR3 and emission-line galaxies in the 
present sample. \citet{G09} have found that galaxies with such strong ongoing 
star formation tend to be shifted to
brighter magnitudes as compared to galaxies in a more quiescent stage.
Using the equivalent width EW(H$\beta$) 
of the H$\beta$ emission line as a measure of the relative
contribution of star forming regions to the total luminosity of the galaxy,
they demonstrated that the galaxies with high EW(H$\beta$) $>$ 100$\AA$\ are 
brighter in the $L-Z$ diagram as compared to the galaxies
with EW(H$\beta$) $<$ 20$\AA$.

Thus, the XMD galaxies (asterisks, open and filled circles) 
do not follow the relations shown by the solid, dashed, and dot-dashed lines. 
They lie in a horizontal band defined by 7.1 $<$ 12 + log O/H $<$ 7.9, 
and their metallicities appear 
to be independent of luminosity over a wide range of galaxy absolute 
brightnesses $-$19 $<$ $M_g$, $M_B$ $<$ $-$13. 
The fact that our emission-line galaxies 
all have relatively low metallicities is not surprising: 
it is a result of the way they have been selected, by considering  
certain emission line flux ratios lying in certain ranges, as described 
before. What may be more surprising is the very large 
range of absolute magnitudes $-$17 $<$ $M_g$, $M_B$ $<$ $-$11 mag 
for the galaxies with the lowest metallicities 
(12 + log O/H $<$ 7.5) in our sample.  
As examples of extreme outliers, we have indicated by stars 
the location of three of the most metal-deficient galaxies known, 
SBS 0335$-$052W, and SBS 0335$-$052E and I Zw 18.
In particular, for its oxygen abundance, the galaxy SBS 0335$-$052E
is some 5.5 mag brighter than the luminosity 
expected from the $L-Z$ relation by \cite{S89} (solid line). 
This high luminosity is due to the presence of 
several luminous super-star clusters in SBS 0335$-$052E \citep{TIL97}.    
\citet{EC10} have attributed the deviations of XMD emission-line galaxies 
from the $L-Z$ relation  
to a combination of a high H {\sc i} gas content (they have processed less gas 
into stars than other galaxies)
and of a more uniform mixing of metals in their ISM due to tidal interactions.
Thus, the use of the luminosity-metallicity relation 
to pick out XMD galaxies, especially the bright ones, is 
not as reliable  as our selection method
based on the relative fluxes of emission lines.

In Figure \ref{fig5}b, we show the relation between the H$\beta$ luminosity 
$L$(H$\beta$) and the oxygen abundance for the same galaxies as in Figure 
\ref{fig5}a. The upper abscissa shows the corresponding star formation
rate (SFR), derived from the H$\alpha$ luminosity and the relation
SFR = 7.9$\times$10$^{-42}$$L$(H$\alpha$) of \citet{K98}, where
SFR is in $M_\odot$ yr$^{-1}$ and $L$(H$\alpha$) is in erg s$^{-1}$.
For log$L$(H$\beta$)$<$39.5, it is seen that the emission-line galaxies 
in both the present and \citet{IT07} samples
follow nicely the relation
defined by the SDSS DR3 galaxies, down to the low-luminosity (or low SFR)  
range. There appears to be a flattening for log$L$(H$\beta$)$>$39.5, 
but it is a result of our selection criteria which pick out 
only galaxies with 12 + logO/H $\la$ 7.6 -- 7.8. The joint fit
to the present, the \citet{IT07} and the SDSS DR3 samples gives the
linear regression 
\begin{equation}
{\rm 12+logO/H} = 0.214\times \log L({\rm H}\beta) - 0.510,
\end{equation}
where $L$(H$\beta$) is in erg s$^{-1}$. 

From inspection
of Fig. \ref{fig5}b, we conclude that the 
most metal-deficient H {\sc ii} regions,
those with 
12 + logO/H $\leq$ 7.35, are found among galaxies with the lowest 
H$\beta$ luminosities, those with log $L$(H$\beta$) $\la$ 37.5. Given 
that a single O5V star produces a H$\beta$
luminosity of 4.8$\times$10$^{36}$ erg s$^{-1}$ (or log $L$(O5V)= 36.68) 
\citep{L90}, this means that only one or a few
O stars are sufficient to ionise the ISM in these XMD
emission-line galaxies with low $L$(H$\beta$).
 Thus, a good recipe for finding XMD emission-line galaxies would be to 
search for galaxies with very low 
 H$\beta$ luminosities, those with log $L$(H$\beta$) $\la$ 37.5.
However, this is a very challenging task since in these galaxies, the emission
lines would be intrinsically weak, and only very nearby XMD emission-line
galaxies would be apparently bright enough for reliable 
abundance determination. There are however a few bright galaxies that are 
outliers, including three well-known XMD BCDs with 12+logO/H $<$ 7.3, in 
Fig. \ref{fig5}b: SBS 0335$-$052W, SBS 0335$-$052E and I Zw 18. 
They are bright because there are presently undergoing an intense 
starburst with a high SFR. This starburst produces copious amounts 
of ionising photons resulting in high-excitation H {\sc ii} regions.
   
In summary, our attempts to find I Zw 18-like or SBS 0335$-$052-like objects 
by searching the entire SDSS spectroscopic survey, containing $\sim$ 10$^6$ 
galaxy spectra, have not been a resounding success. No such galaxy was found 
in the SDSS, except for I Zw 18 itself.
The other two galaxies, SBS 0335$-$052W and SBS 0335$-$052E, are outside
the sky region covered by the SDSS. 
Although the number of XMD galaxies with oxygen abundances 
near the metallicity floor has grown, none
of the newly discovered ones listed in Table \ref{tab10} is both as bright in 
apparent magnitude and contains high-excitation H {\sc ii} regions as the two 
well-known prototype XMD BCDs I Zw 18 and SBS 0335--052E,
discovered respectively by \citet{SS70} and \citet{I90}.
Because of their brightness and their high-excitation H {\sc ii} regions, 
these objects have allowed high-sensitivity and high spatial resolution 
multiwavelength studies of their 
star formation and chemical evolution properties that none of the 
newly discovered XMD emission-line galaxies will permit 
\citep[see ][ for a review]{T08}. After more than 40 years of search for 
the most metal-poor galaxies, I Zw 18 and SBS 0335--052E still remain 
the most interesting objects among XMD emission-line galaxies.

\section{XMD emission-line galaxies and the emission-line diagnostic 
diagram\label{S5}}

\setcounter{figure}{5}


\begin{figure}
\hspace*{0.0cm}\psfig{figure=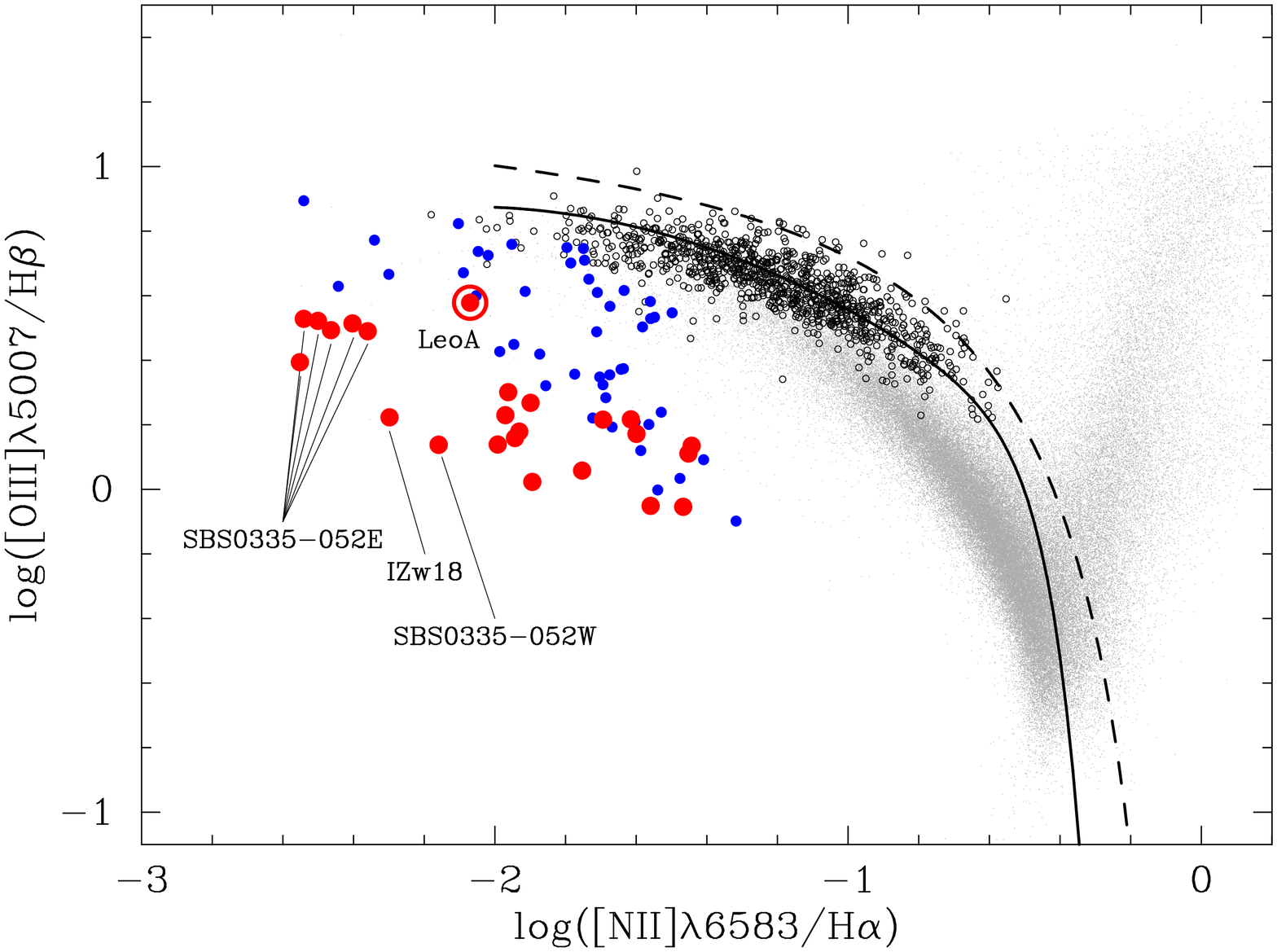,angle=-90,width=8.8cm,clip=}
\caption{The Baldwin-Phillips-Terlevich diagram \citep{BPT81} for emission-line
galaxies. The emission-line galaxies observed in this paper are shown by 
filled blue circles and the most-metal deficient star-forming galaxies known 
(Table \ref{tab10}) are shown by red filled
circles, including observations of several knots of star formation
in SBS 0335--052E. The Local Group dwarf irregular 
galaxy Leo A is encircled. Also, plotted 
are the 100,000 emission-line galaxies from SDSS DR7
(cloud of grey dots) and a sample of 803 luminous compact galaxies (LCGs, open 
black circles) \citep{I11}. The dashed line from \citet{K03} and the solid
line from \citet{S06} separate star-forming galaxies from active galactic 
nuclei. The latter lie in the right wing of the ``butterfly''. }
\label{fig6}
\end{figure}

Finally, in Fig. \ref{fig6} we show the location of the most metal-poor 
galaxies in the Baldwin-Phillips-Terlevich (BPT) diagnostic 
diagram \citep{BPT81} for emission-line galaxies. The cloud of grey dots 
represents the location of the $\sim$ 100,000 SDSS DR7 galaxies. The dashed
and solid lines \citep{K03,S06} separate star-forming galaxies (to the left
of the lines) from active galactic nuclei (to the right of the lines).
Open black circles are luminous compact galaxies (LCGs) with high SFRs 
of $\sim$ 1 - 60 $M_\odot$ yr$^{-1}$ which occupy 
the upper tail of the star-forming galaxy region. LCGs were studied
in detail by \citet{I11} who found that their oxygen abundance 
distribution peaks at 12 + log O/H
of $\sim$ 8.0. The emission-line
galaxies observed and discussed in the present paper are shown by 
filled blue circles. They are located below the region where lie the bulk  
of the star-forming galaxies, a sign of their less advanced chemical 
evolution and of their lower metallicities. The most metal-deficient 
galaxies known, listed in Table \ref{tab10}, are shown by red filled circles. 
We have plotted all 
observations of these galaxies that we could find in the 
literature, including multiple observations of SBS 0335--052E. 
It is evident from Fig. \ref{fig6} that the lowest-metallicity 
emission-line galaxies 
occupy a region of the BPT diagram, which has been left 
vacant before by the known star-forming galaxy population. 
The lowest-metallicity galaxies 
have considerably smaller [O {\sc iii}]/H$\beta$ ratios and generally 
smaller [N {\sc ii}]/H$\alpha$ ratios. The Local Group dwarf irregular 
galaxy Leo A (encircled in Fig. \ref{fig6}) 
has the highest [O {\sc iii}]/H$\beta$ ratio among the known 
lowest-metallicity emission-line galaxies. We note that  
the oxygen abundance for Leo A has been derived from the spectrum of a 
planetary nebula, not of a H {\sc ii} region \citep{S89,V06}.     
The fact that this region in the BPT diagram was left empty before    
suggests that the population of the most metal-poor emission-line
galaxies in the local Universe is extremely small. 
This is born out by our rough search statistics.
Out of one million SDSS spectra, we have isolated about 13000 
emission-line objects with a detected [O {\sc iii}]$\lambda$4363, and out 
of these there are $\sim$ 15 galaxies with 12 + log O/H $\leq$ 7.35, 
so the fraction of the most metal-poor emission-line galaxies in the 
local universe is $\sim$0.1\%.

\section{Conclusions \label{S6}}

We present spectroscopic observations with the 6.5m MMT 
and the 3.5m APO telescope of
a sample of 69 H {\sc ii} regions in 42 dwarf emission-line
galaxies. These galaxies were selected from the Data Release 7 (DR7)
of the Sloan Digital Sky Survey (SDSS) using mainly two criteria:   
[O {\sc iii}]$\lambda$4959/H$\beta$ $\la$ 1 and 
[N {\sc ii}]$\lambda$6583/H$\beta$ $\la$ 0.1.
These spectral properties select out extremely low-metallicity
galaxies,  with oxygen
abundances comparable to those of the two prototype 
extremely metal-deficient (XMD) 
emission-line galaxies SBS 0335--052E and I~Zw~18. 

Our results are as follows:

1. We find that 14 H {\sc ii} regions in 7 emission-line galaxies
have oxygen abundances
12 + log O/H $\leq$ 7.35. Among them, the 2 galaxies
J0113+0052 and DDO 68 have been observed previously.
We confirm the very low oxygen abundances for J0113+0052 found
by \citet{I06b} and for DDO 68 found by \citet{P05} and \citet{IT07}.
In addition, we find 5 more galaxies with 12 + log O/H $\leq$ 7.35. 
Our data set, when combined with previous studies, 
results in a total of 17 known XMD emission-line galaxies with 
12 + log O/H $\leq$ 7.35 (Table \ref{tab10}). 
Among all the emission-line galaxies that have been studied, 
there is no object that has been discovered with 12 + logO/H $<$ 6.9,
although the selection criteria do not forbid such an object. 
This appears to suggest the existence 
of an oxygen abundance floor at that level, and supports
the idea that the matter from which dwarf emission-line galaxies
formed was pre-enriched to a level 12 + log O/H $\sim$ 6.9  
\citep[e.g., ][]{T05}. 

2. We have examined the absolute magnitude - metallicity diagram
for our emission-line galaxies. We find that these form an horizontal 
band where the metallicity appears to be nearly independent of luminosity over 
a wide range ($\sim$ 8 magnitudes) of absolute magnitudes, corresponding to
a luminosity range of $\sim$ 1500. 
The inclusion of XMD emission-line galaxies in a sample of dwarf 
star-forming galaxies would make the slope of the absolute magnitude - 
metallicity relation shallower as compared to samples of more quiescent 
dwarf irregular galaxies, not including XMD emission-line galaxies, 
as those of \citet{S89} and \citet{RM95}. On the other hand, we find a tight 
relation between the H$\beta$ luminosity and the 
metallicity for the combined sample of XMD and SDSS emission-line galaxies,
where the luminosities change by a factor of $\sim$ 10$^5$.

3. Because of the large dispersion about the fit to the absolute 
magnitude - metallicity relation, the use of the latter is not a good way to 
select XMD emission-line galaxies.
It is more efficient to select them on the basis of 
flux ratios of nebular [O~{\sc iii}] and [N~{\sc ii}] relative 
to H$\beta$ emission lines, and by using the H$\beta$ luminosity -
metallicity relation. 
We find that more than $\sim$ 1/3 of the studied H {\sc ii} regions with 
log $L$(H$\beta$) $\la$ 37.5 are XMD galaxies with 12 + logO/H $\leq$ 7.35. 
On the other hand, we have failed to find more bright extremely metal-poor 
emission-line galaxies in the complete SDSS spectroscopic survey, similar to 
the prototype BCDs SBS 0335$-$052E and I Zw 18.

\acknowledgements

This research has been supported by NSF grant AST-02-05785 
and NASA contract 1263707.
Y.I.I. thanks the hospitality of the Astronomy Department of 
the University of Virginia, USA. Y.I.I. and N.G.G. thank 
the hospitality of the Max-Planck 
Institute for Radioastronomy, Bonn, Germany. T.X.T. thanks the 
hospitality of the Institut d'Astrophysique de Paris.   
    Funding for the Sloan Digital Sky Survey (SDSS) and SDSS-II has been 
provided by the Alfred P. Sloan Foundation, the Participating Institutions, 
the National Science Foundation, the U.S. Department of Energy, the National 
Aeronautics and Space Administration, the Japanese Monbukagakusho, and the 
Max Planck Society, and the Higher Education Funding Council for England. 
 

\Online





\setcounter{table}{1}

\begin{longtable}{lrrccc}
  \caption{Journal of Observations}
\label{tab2} \\ \hline \hline
\endfirsthead
\caption{---Continued.}\\
\hline\hline
Name&Date&Exposure&Position angle&Airmass&Slit \\
&&(s)&(deg)&&(arcsec) \\ \hline
\endhead
\hline
\endfoot
Name&Date&Exposure&Position angle&Airmass&Slit \\
&&(s)&(deg)&&(arcsec) \\ \hline
\multicolumn{6}{c}{a) 3.5m APO observations} \\ \hline
J0113$+$0052No.1&28 Oct. 2008&2700& $+$62.8&1.41&1.5 \\
J0113$+$0052No.2&28 Oct. 2008&2700& $+$62.8&1.41&1.5 \\
J0113$+$0052No.3&28 Oct. 2008&2700& $+$62.8&1.41&1.5 \\
J0113$+$0052No.4&28 Oct. 2008&2700& $+$11.9&1.75&1.5 \\
J0138$+$0020    &07 Feb. 2008&1800& $+$72.3&1.50&1.5 \\
\end{longtable}

\setcounter{table}{2}

\begin{longtable}{lrrrrr}
  \caption{Emission line intensities in 3.5m APO spectra}
\label{tab3} \\ \hline \hline
\endfirsthead
\caption{---{\sl Continued.}}\\
\hline\hline
  {Ion}
  &{$I$($\lambda$)/$I$(H$\beta$)}
  &{$I$($\lambda$)/$I$(H$\beta$)}
  &{$I$($\lambda$)/$I$(H$\beta$)}
  &{$I$($\lambda$)/$I$(H$\beta$)}
  &{$I$($\lambda$)/$I$(H$\beta$)} \\
  \cline{2-6}
  {}&\multicolumn{5}{c}{\sc Galaxy} \\  \hline    
\endhead
\hline
\endfoot
  {Ion}
  &{$I$($\lambda$)/$I$(H$\beta$)}
  &{$I$($\lambda$)/$I$(H$\beta$)}
  &{$I$($\lambda$)/$I$(H$\beta$)}
  &{$I$($\lambda$)/$I$(H$\beta$)}
  &{$I$($\lambda$)/$I$(H$\beta$)} \\
  \cline{2-6}
  {}&\multicolumn{5}{c}{\sc Galaxy} \\  \hline    
  {}&
 {    J0113+0052No.1}&{    J0113+0052No.2}&
 {    J0113+0052No.3}&{    J0113+0052No.4}&
 {    J0138+0020    } \\ \hline
3727 [O {\sc ii}]                 &  82.51 $\pm$   7.09 & 198.72 $\pm$  13.11 & 152.91 $\pm$  10.27 &  84.76 $\pm$   5.96 & 224.87 $\pm$  32.55 \\
3868 [Ne {\sc iii}]               &  23.73 $\pm$   3.41 &   ...~~~~~~~~       &   ...~~~~~~~~       &   ...~~~~~~~~       &   ...~~~~~~~~       \\
3889 He {\sc i} + H8              &  23.86 $\pm$   4.79 &   ...~~~~~~~~       &   ...~~~~~~~~       &   ...~~~~~~~~       &   ...~~~~~~~~       \\
3968 [Ne {\sc iii}] + H7          &  20.61 $\pm$   4.08 &   ...~~~~~~~~       &   ...~~~~~~~~       &   ...~~~~~~~~       &   ...~~~~~~~~       \\
4101 H$\delta$                    &  26.01 $\pm$   3.87 &  31.54 $\pm$   3.87 &   ...~~~~~~~~       &   ...~~~~~~~~       &   ...~~~~~~~~       \\
\end{longtable}

\setcounter{table}{3}

\begin{longtable}{lrrrrr}
  \caption{Emission line intensities in low-resolution MMT spectra}
\label{tab4} \\ \hline \hline
\endfirsthead
\caption{---{\sl Continued.}}\\
\hline\hline
  {Ion}
  &{$I$($\lambda$)/$I$(H$\beta$)}
  &{$I$($\lambda$)/$I$(H$\beta$)}
  &{$I$($\lambda$)/$I$(H$\beta$)}
  &{$I$($\lambda$)/$I$(H$\beta$)}
  &{$I$($\lambda$)/$I$(H$\beta$)} \\
  \cline{2-6}
  {}&\multicolumn{5}{c}{\sc Galaxy} \\  \hline    
\endhead
\hline
\endfoot
  {Ion}
  &{$I$($\lambda$)/$I$(H$\beta$)}
  &{$I$($\lambda$)/$I$(H$\beta$)}
  &{$I$($\lambda$)/$I$(H$\beta$)}
  &{$I$($\lambda$)/$I$(H$\beta$)}
  &{$I$($\lambda$)/$I$(H$\beta$)} \\
  \cline{2-6}
  {}&\multicolumn{5}{c}{\sc Galaxy} \\      \cline{2-6}
  {}&
 {    J1016+3754     }&{    J1016+5823No.1 }&
 {    J1016+5823No.2 }&{    J1016+5823No.3 }&
 {    J1044+0353     } \\ \hline
3727 [O {\sc ii}]                 &  62.50 $\pm$   1.26 & 114.63 $\pm$   2.03 & 134.53 $\pm$   5.63 & 104.00 $\pm$   2.04 &  32.64 $\pm$   0.70 \\
3750 H12                          &   3.14 $\pm$   0.76 &   2.61 $\pm$   0.63 &   ...~~~~~~~~       &   3.00 $\pm$   0.83 &   3.07 $\pm$   0.37 \\
3771 H11                          &   3.81 $\pm$   0.82 &   2.93 $\pm$   0.58 &   ...~~~~~~~~       &   3.30 $\pm$   0.79 &   3.95 $\pm$   0.38 \\
3798 H10                          &   4.19 $\pm$   0.67 &   4.07 $\pm$   0.52 &   ...~~~~~~~~       &   4.44 $\pm$   0.76 &   5.38 $\pm$   0.36 \\
3820 He {\sc i}                   &   ...~~~~~~~~       &   1.09 $\pm$   0.35 &   ...~~~~~~~~       &   ...~~~~~~~~       &   ...~~~~~~~~       \\
\end{longtable}

\setcounter{table}{4}

\begin{longtable}{lrrrrr}
  \caption{Emission line intensities in medium-resolution MMT spectra}
\label{tab5} \\ \hline \hline
\endfirsthead
\caption{---{\sl Continued.}}\\
\hline\hline
  {Ion}
  &{$I$($\lambda$)/$I$(H$\beta$)}
  &{$I$($\lambda$)/$I$(H$\beta$)}
  &{$I$($\lambda$)/$I$(H$\beta$)}
  &{$I$($\lambda$)/$I$(H$\beta$)}
  &{$I$($\lambda$)/$I$(H$\beta$)} \\
  \cline{2-6}
  {}&\multicolumn{5}{c}{\sc Galaxy} \\  \hline    
\endhead
\hline
\endfoot
  {Ion}
  &{$I$($\lambda$)/$I$(H$\beta$)}
  &{$I$($\lambda$)/$I$(H$\beta$)}
  &{$I$($\lambda$)/$I$(H$\beta$)}
  &{$I$($\lambda$)/$I$(H$\beta$)}
  &{$I$($\lambda$)/$I$(H$\beta$)} \\
  \cline{2-6}
  {}&\multicolumn{5}{c}{\sc Galaxy} \\      \cline{2-6}
  {}&
 {    J0810+1837     }&{    J0833+2508 No.1}&
 {    J0833+2508 No.2}&{    J0931+2717 No.1}&
 {    J0931+2717 No.2} \\ \hline
3727 [O {\sc ii}]                 & 333.47 $\pm$  19.68 & 215.86 $\pm$   4.70 & 247.67 $\pm$  11.39 & 222.07 $\pm$  10.10 & 239.73 $\pm$  10.47 \\
3835 H9                           &   ...~~~~~~~~       &   9.13 $\pm$   1.41 &   ...~~~~~~~~       &   ...~~~~~~~~       &   ...~~~~~~~~       \\
3868 [Ne {\sc iii}]               &  26.50 $\pm$   4.32 &  22.93 $\pm$   1.15 &  31.27 $\pm$   4.11 &  12.55 $\pm$   2.46 &  13.68 $\pm$   2.84 \\
3889 He {\sc i} + H8              &   ...~~~~~~~~       &  20.06 $\pm$   1.38 &  21.80 $\pm$   4.84 &   ...~~~~~~~~       &  23.72 $\pm$   7.27 \\
3968 [Ne {\sc iii}] + H7          &   ...~~~~~~~~       &  21.04 $\pm$   1.33 &   ...~~~~~~~~       &   ...~~~~~~~~       &  18.01 $\pm$   6.66 \\
\end{longtable}

\setcounter{table}{5}

\begin{longtable}{lrrrrr}
  \caption{Ionic and total heavy element abundances derived from 3.5m APO
spectra}
\label{tab6} \\ \hline \hline
\endfirsthead
\caption{---{\sl Continued.}}\\
\hline\hline
  {}&\multicolumn{5}{c}{Galaxy}\\       \cline{2-6}
\endhead
\hline
\endfoot
  {}&\multicolumn{5}{c}{Galaxy}\\       \cline{2-6}
  {Property}&
\multicolumn{1}{c}{    J0113+0052No.1$^a$} & \multicolumn{1}{c}{    J0113+0052No.2$^a$} &
\multicolumn{1}{c}{    J0113+0052No.3$^a$} & \multicolumn{1}{c}{    J0113+0052No.4$^a$} &
\multicolumn{1}{c}{    J0138+0020$^b$    } \\ \hline
  $T_{\rm e}$(O {\sc iii}) (K) &
$22427 \pm3086 $ & $19102  \pm1032 $ &
$26186  \pm3779 $ & $19007  \pm1010 $ &
$17131  \pm1110 $  \\

  $T_{\rm e}$(O {\sc ii}) (K) &
$16379  \pm1904 $ & $16103  \pm 778 $ &
$15387  \pm1375 $ & $16079  \pm 764 $ &
$15427  \pm1343 $  \\

  $T_{\rm e}$(S {\sc iii}) (K) &
$20314  \pm2561 $ & $17555  \pm 857 $ &
$23435  \pm3136 $ & $17476  \pm 838 $ &
$15919  \pm 922 $  \\

  \\
  O$^+$/H$^+$ ($\times$10$^4$) &
$ 0.058 \pm 0.017$ & $ 0.146 \pm 0.020$ &
$ 0.128 \pm 0.030$ & $ 0.062 \pm 0.008$ &
$ 0.187 \pm 0.049$ \\
  O$^{++}$/H$^+$ ($\times$10$^4$) &
$ 0.085 \pm 0.025$ & $ 0.063 \pm 0.008$ &
$ 0.071 \pm 0.021$ & $ 0.121 \pm 0.015$ &
$ 0.144 \pm 0.025$ \\
\end{longtable}

\setcounter{table}{6}

\begin{longtable}{lrrrrr}
  \caption{Ionic and total heavy element abundances derived from the
low-resolution MMT spectra}
\label{tab7} \\ \hline \hline
\endfirsthead
\caption{---{\sl Continued.}}\\
\hline\hline
  {}&\multicolumn{5}{c}{Galaxy}\\       \cline{2-6}
\endhead
\hline
\endfoot
  {}&\multicolumn{5}{c}{Galaxy}\\       \cline{2-6}
  {Property}&
\multicolumn{1}{c}{J1016+3754$^a$     } & \multicolumn{1}{c}{J1016+5823No.1$^a$ } &
\multicolumn{1}{c}{J1016+5823No.2$^a$ } & \multicolumn{1}{c}{J1016+5823No.3$^a$ } &
\multicolumn{1}{c}{J1044+0353$^a$     } \\ \hline
  $T_{\rm e}$(O {\sc iii}) (K) &
$17167  \pm 214 $ & $15410  \pm 205 $ &
$15170  \pm1093 $ & $16960  \pm 265 $ &
$19881  \pm 256 $  \\

  $T_{\rm e}$(O {\sc ii}) (K) &
$15443  \pm 174 $ & $14526  \pm 177 $ &
$14377  \pm 947 $ & $15351  \pm 217 $ &
$16265  \pm 186 $  \\

  $T_{\rm e}$(S {\sc iii}) (K) &
$15949  \pm 178 $ & $14968  \pm 170 $ &
$14774  \pm 907 $ & $15776  \pm 220 $ &
$18201  \pm 212 $  \\

  \\
  O$^+$/H$^+$ ($\times$10$^4$) &
$ 0.052 \pm 0.002$ & $ 0.115 \pm 0.004$ &
$ 0.139 \pm 0.026$ & $ 0.088 \pm 0.004$ &
$ 0.024 \pm 0.001$ \\
  O$^{++}$/H$^+$ ($\times$10$^4$) &
$ 0.373 \pm 0.012$ & $ 0.422 \pm 0.015$ &
$ 0.253 \pm 0.047$ & $ 0.327 \pm 0.013$ &
$ 0.247 \pm 0.008$ \\
\end{longtable}

\setcounter{table}{7}

\newpage

  \begin{table}
  \caption{Ionic and total heavy element abundances derived from the
high-resolution MMT spectra \label{tab8}}
\begin{tabular}{lrrrrr} \hline \hline
  {}&\multicolumn{5}{c}{Galaxy}\\       \cline{2-6}
  {Property}&
\multicolumn{1}{c}{     J0810+1837$^a$     } & \multicolumn{1}{c}{     J0833+2508 No.1$^a$} &
\multicolumn{1}{c}{     J0833+2508 No.2$^a$} & \multicolumn{1}{c}{     J0931+2717 No.1$^b$} &
\multicolumn{1}{c}{     J0931+2717 No.2$^b$} \\ \hline
  $T_{\rm e}$(O {\sc iii}) (K) &
$14517  \pm3643 $ & $15635  \pm 820 $ &
$21239  \pm4080 $ & $17809  \pm1015 $ &
$17147  \pm1014 $  \\

  $T_{\rm e}$(O {\sc ii}) (K) &
$13945  \pm3212 $ & $14661  \pm 702 $ &
$16405  \pm2740 $ & $15702  \pm1241 $ &
$15434  \pm1226 $  \\

  \\
  O$^+$/H$^+$ ($\times$10$^4$) &
$ 0.381 \pm 0.247$ & $ 0.210 \pm 0.027$ &
$ 0.172 \pm 0.069$ & $ 0.175 \pm 0.035$ &
$ 0.199 \pm 0.041$ \\
  O$^{++}$/H$^+$ ($\times$10$^4$) &
$ 0.292 \pm 0.193$ & $ 0.281 \pm 0.038$ &
$ 0.100 \pm 0.042$ & $ 0.106 \pm 0.015$ &
$ 0.134 \pm 0.020$ \\
  O$^{+++}$/H$^+$ ($\times$10$^6$) &
$ ...~~~~~~~~~~$ & $ ...~~~~~~~~~~$ &
$ ...~~~~~~~~~~$ & $ ...~~~~~~~~~~$ &
$ ...~~~~~~~~~~$ \\
\hline
  \end{tabular}

$^a$direct method.

$^b$semi-empirical method.
  \end{table}




\setcounter{figure}{0}

\begin{figure*}
\hspace*{0.0cm}\psfig{figure=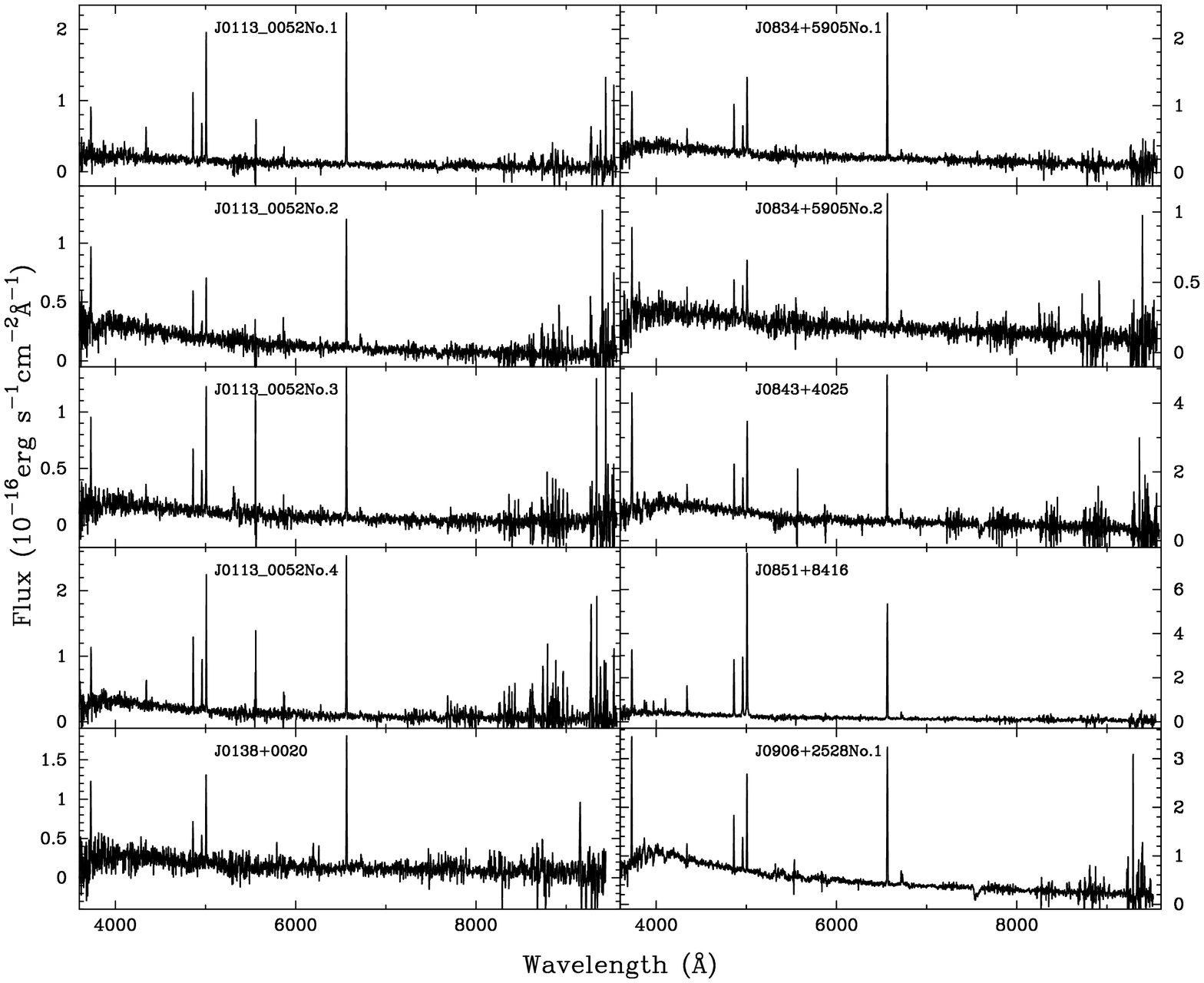,angle=-90,width=19.cm,clip=}
\caption{3.5m APO spectra of galaxies. }
\label{fig1}
\end{figure*}

\setcounter{figure}{0}

\begin{figure*}
\hspace*{0.0cm}\psfig{figure=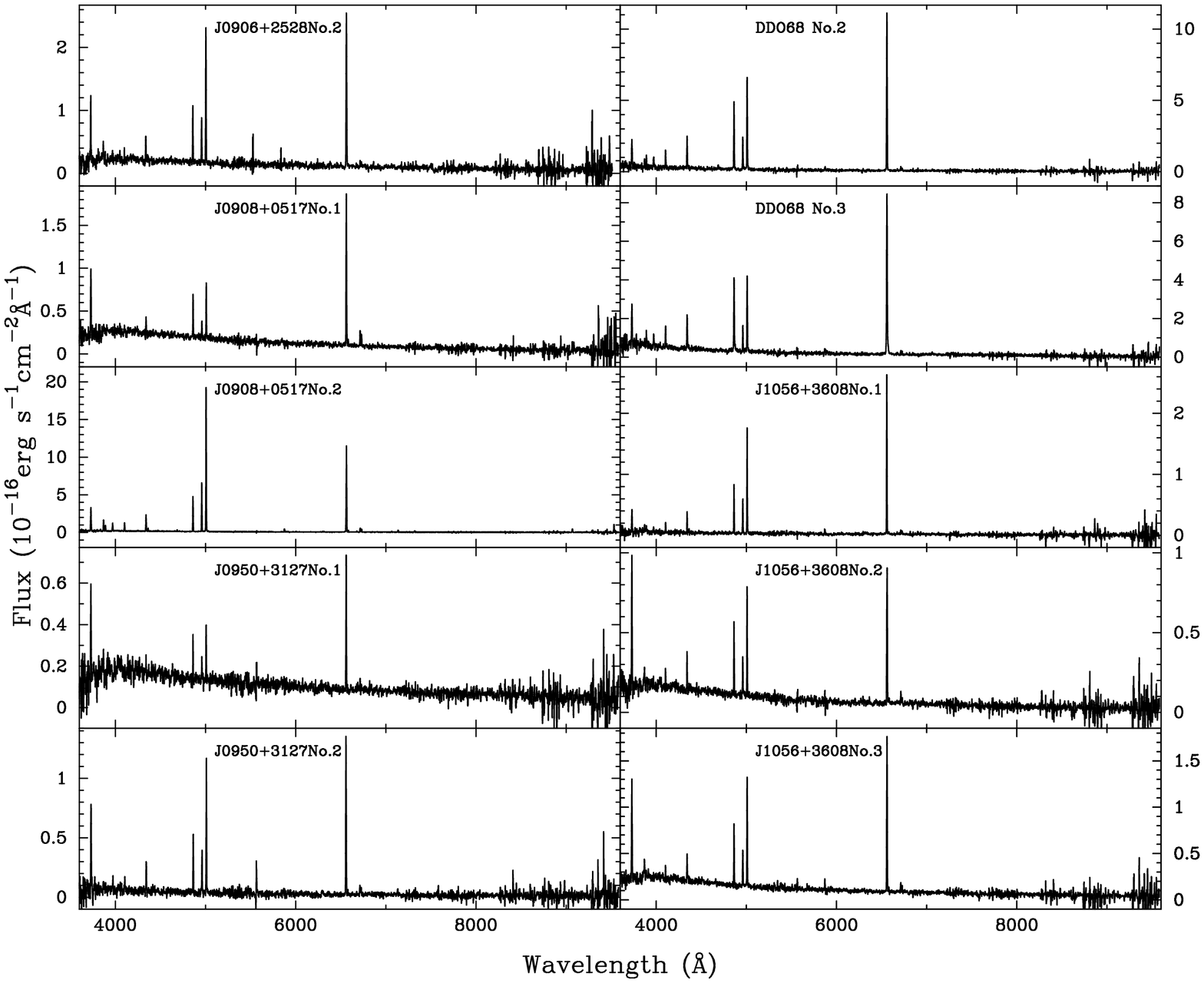,angle=-90,width=19.cm,clip=}
\caption{---{\sl Continued.} }
\end{figure*}

\setcounter{figure}{0}

\begin{figure*}
\hspace*{0.0cm}\psfig{figure=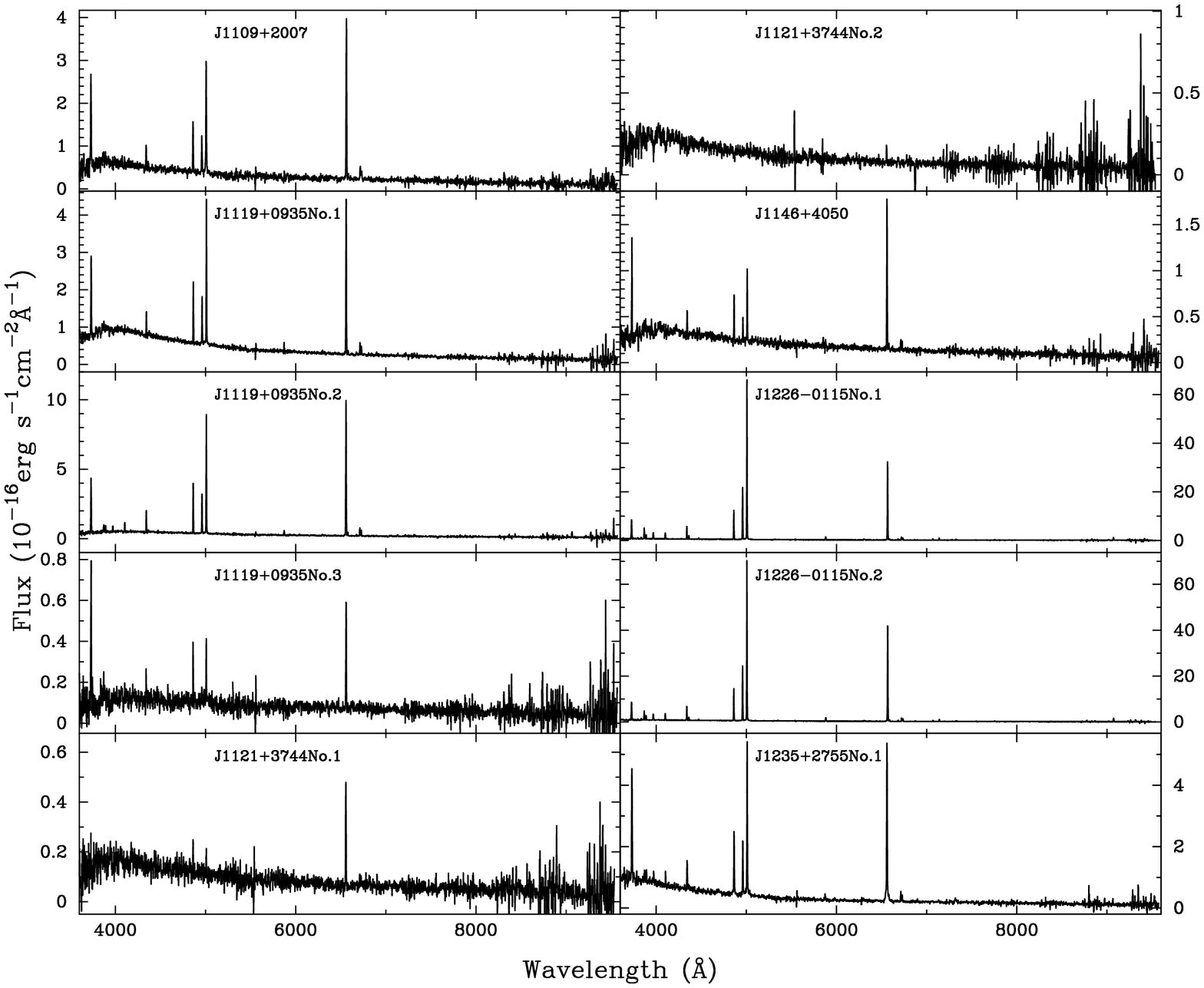,angle=-90,width=19.cm,clip=}
\caption{---{\sl Continued.} }
\end{figure*}

\setcounter{figure}{0}

\begin{figure*}
\hspace*{0.0cm}\psfig{figure=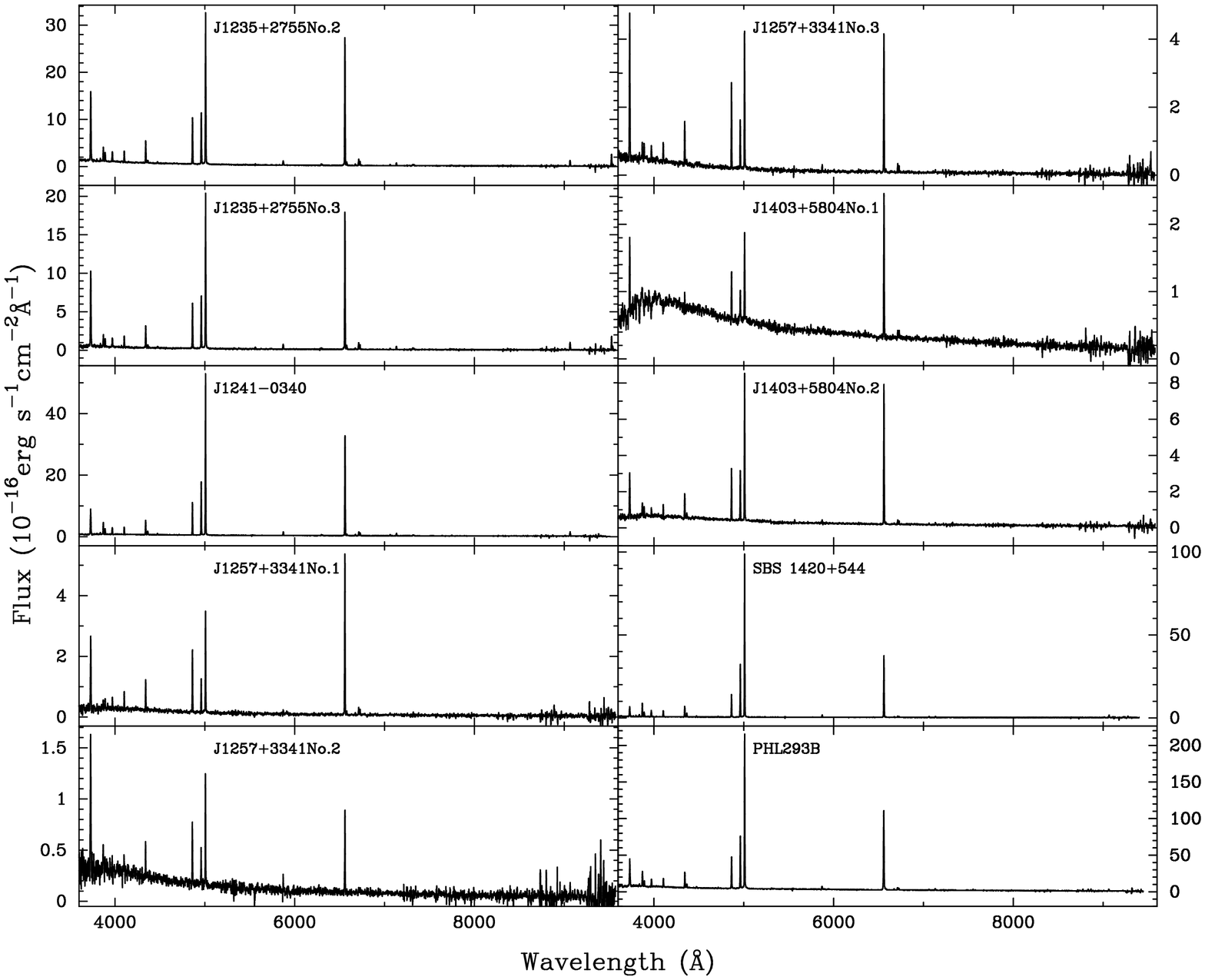,angle=-90,width=19.cm,clip=}
\caption{---{\sl Continued.} }
\end{figure*}

\begin{figure*}
\hspace*{0.0cm}\psfig{figure=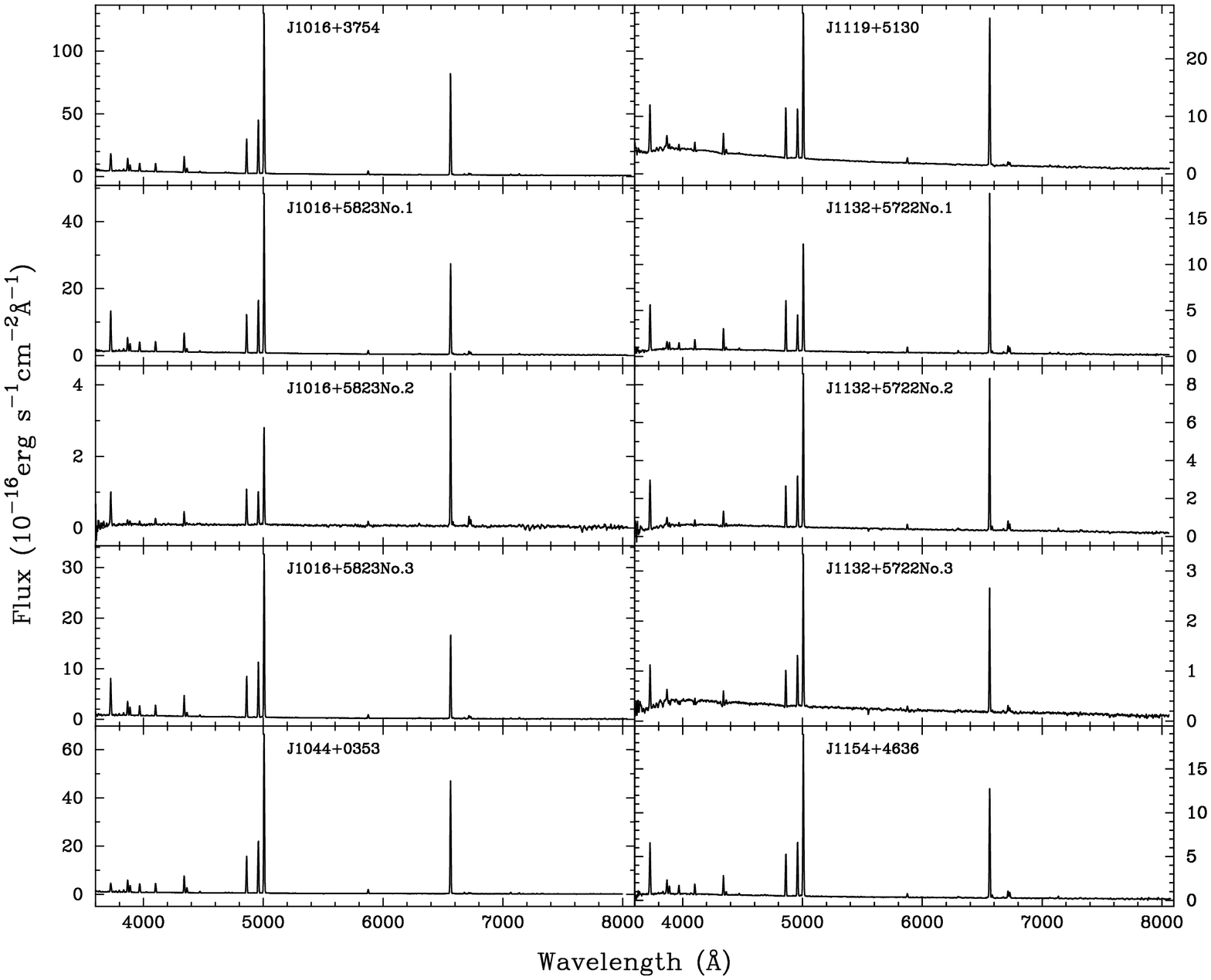,angle=-90,width=19.cm,clip=}
\caption{Low-resolution MMT spectra. }
\label{fig2}
\end{figure*}

\setcounter{figure}{1}

\begin{figure*}
\hspace*{0.0cm}\psfig{figure=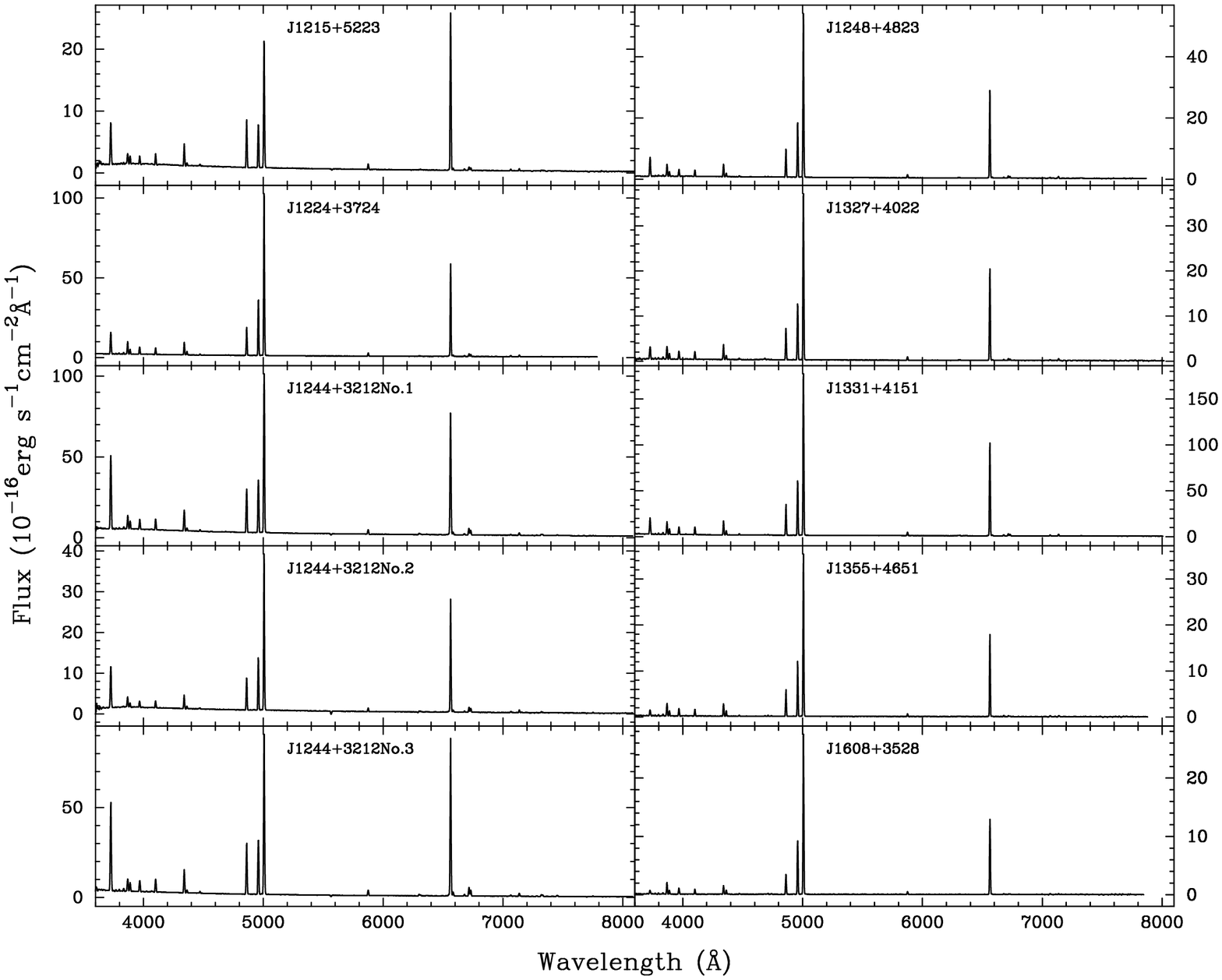,angle=-90,width=19.cm,clip=}
\caption{---{\sl Continued.} }
\end{figure*}

\begin{figure*}
\hspace*{0.0cm}\psfig{figure=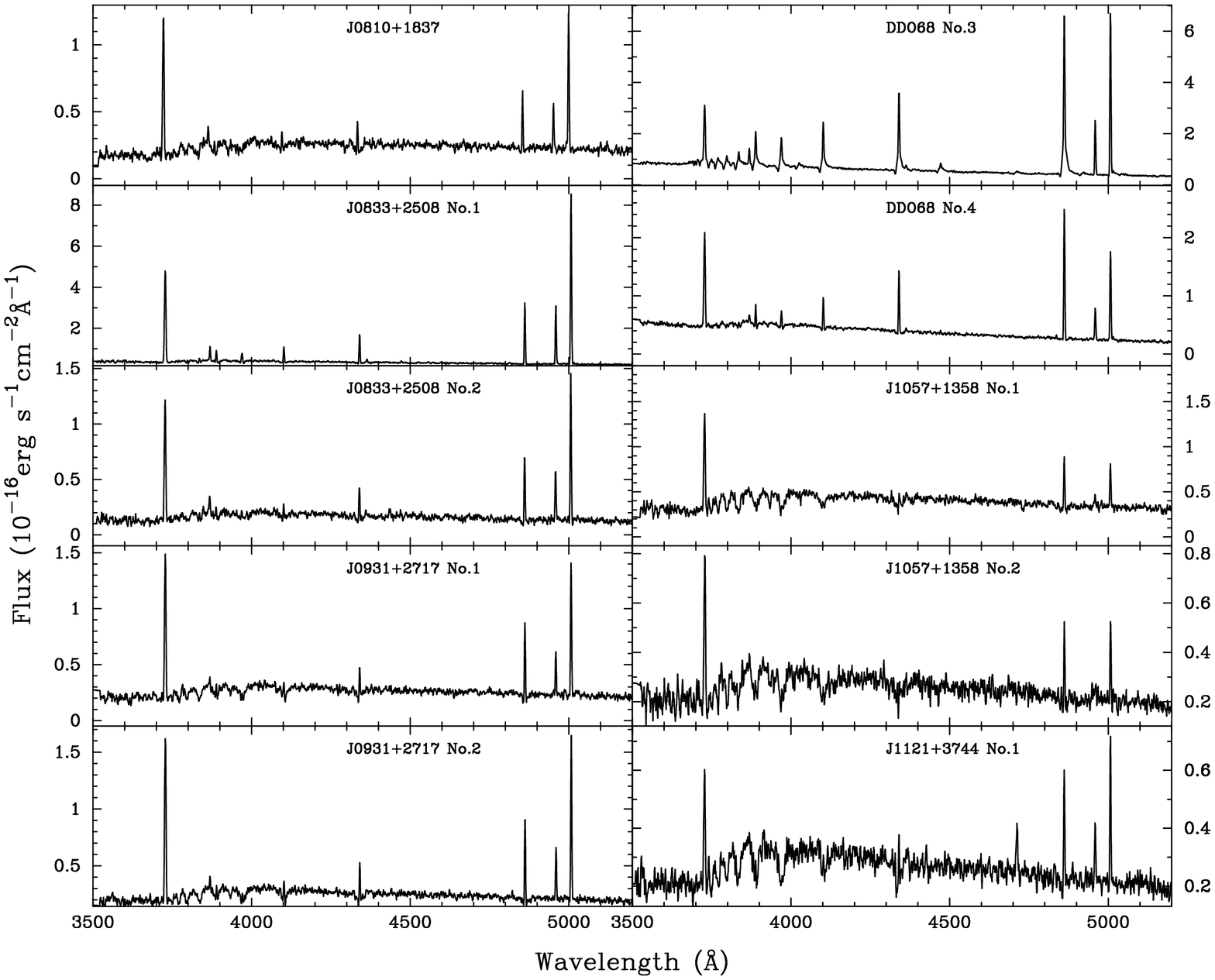,angle=-90,width=19.cm,clip=}
\caption{Medium-resolution MMT spectra. }
\label{fig3}
\end{figure*}

\setcounter{figure}{2}


\begin{figure*}
\hspace*{0.0cm}\psfig{figure=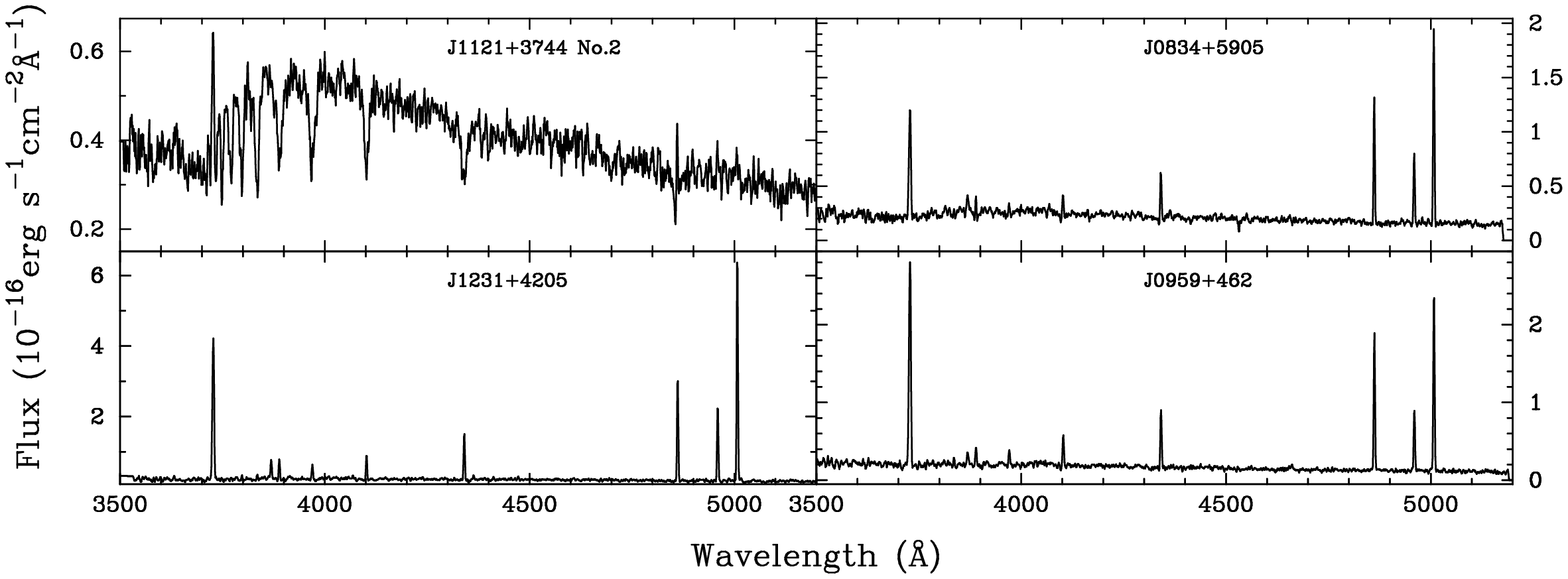,angle=-90,width=19.cm,clip=}
\caption{---{\sl Continued.} }
\end{figure*}


\end{document}